%% file: 2022_dragondoom_to_dragonstar.tex
\documentclass[compsoc, conference, a4paper, 10pt, times]{./IEEEtran}
\usepackage{cite}
\usepackage{amsfonts}
\usepackage[cmex10]{amsmath} 
\usepackage{array}
\usepackage{graphicx}
\usepackage{subcaption}
\usepackage{pgfplots}
\usepackage{booktabs}
\usepackage{multirow}
\usepackage{listings}
\usepackage{verbatimbox}
\usepackage{hyperref}
\usepackage{threeparttable}
\usepackage{tablefootnote}
\usepackage{paralist}

\definecolor{bblue}{rgb}{0.1,0.51,1}
\definecolor{rred}{rgb}{0.83,0.07,0.35}

\definecolor{linkcolor}{rgb}{0.65,0,0}
\definecolor{citecolor}{rgb}{0,0.65,0}
\definecolor{urlcolor}{rgb}{0,0,0.65}
\hypersetup{colorlinks=true, linkcolor=linkcolor, urlcolor=urlcolor, citecolor=citecolor}
\usepackage{url}

\definecolor{dkgreen}{rgb}{0,0.6,0}
\definecolor{gray}{rgb}{0.5,0.5,0.5}
\definecolor{lightgray}{rgb}{0.8,0.8,0.8}
\definecolor{mauve}{rgb}{0.58,0,0.82}
\lstdefinestyle{codesample}{
	basicstyle=\footnotesize,
	frame=lines,
	language=python,
	aboveskip=3mm,
	belowskip=3mm,
	showstringspaces=false,
	columns=flexible,
	numbers=left,
	numberstyle=\footnotesize\color{black},
	numbersep=5pt,
	morekeywords={False, True, uint8\_t, bool},
	keywordstyle=\color{blue},
	commentstyle=\color{dkgreen},
	stringstyle=\color{mauve},
	breaklines=true,
	breakatwhitespace=true,
	tabsize=2,
	captionpos=b,
	belowskip=-1.75 \baselineskip
}
\pgfplotsset{compat=1.16}

\usetikzlibrary{calc,positioning,shapes.misc}

\makeatletter
\AtBeginDocument{\DeclareCaptionSubType{lstlisting}
	\renewcommand{\p@sublstlisting}{\thelstlisting}
	
}

\renewcommand{\paragraph}{%
	\@startsection{paragraph}{4}%
	{\z@}{1ex \@plus 1ex \@minus .2ex}{-0.5em}%
	{\normalfont\normalsize\bfseries}%
}
\makeatother

\usepackage{xcolor}
\input{resources/lstfstar}
\newcommand\maybecolor[1]{\color{#1}}
\definecolor{dkblue}{rgb}{0,0.1,0.5}
\definecolor{dkred}{rgb}{0.6,0,0}

\usepackage{varwidth}
\usepackage{tcolorbox}
\tcbuselibrary{skins,fitting}
\tcbsetforeverylayer{shield externalize}
\definecolor {todoboxcolor}{HTML}{AC0B0B}
\newtcolorbox{todobox}[1]{%
	enhanced,
	skin=enhancedlast jigsaw,
	title={\textcolor{white}{\textbf{TODO: #1}}},
	attach boxed title to top left={xshift=-4mm,yshift=-0.5mm},
	fonttitle=\bfseries\sffamily,
	varwidth boxed title=0.7\linewidth,
	colbacktitle=todoboxcolor,
	boxed title style={
		empty,
		arc=0pt,
		outer arc=0pt,
		boxrule=0pt
	},
	underlay boxed title={
		\fill[todoboxcolor] (title.north west) -- (title.north east)
		-- +(\tcboxedtitleheight-1mm,-\tcboxedtitleheight+1mm)
		-- ([xshift=4mm,yshift=0.5mm]frame.north east) -- +(0mm,-1mm)
		-- (title.south west) -- cycle;
		\fill[todoboxcolor!45!white!50!black] ([yshift=-0.5mm]frame.north west)
		-- +(-0.4,0) -- +(0,-0.3) -- cycle;
		\fill[todoboxcolor!45!white!50!black] ([yshift=-0.5mm]frame.north east)
		-- +(0,-0.3) -- +(0.4,0) -- cycle; 
	},
	colframe=todoboxcolor,%!10,
	colback=todoboxcolor!10,
	top=5pt,
	bottom=5pt,
	boxsep=2pt,
	beforeafter skip=0.5em,
	leftright skip=0em,
}

\newsavebox{\verbsavebox}

\newcommand{\fr}{Flush+Reload }
\newcommand{\eg}{\textit{e.g.}, }
\newcommand{\ie}{\textit{i.e.}, }
\def\rot{\rotatebox}
\newcommand{\aref}[1]{\hyperref[#1]{appendix~\ref*{#1}}}

\graphicspath{{resources/}}

\newcommand{\code}[1]{\texttt{\footnotesize{#1}}}

\begin{document}
%-------------------------------------------------------------------------------

%don't want date printed
\date{}

\title{From Dragondoom to Dragonstar: Side-channel Attacks and Formally Verified Implementation of WPA3 Dragonfly Handshake}
\author{
	\IEEEauthorblockN{Daniel De Almeida Braga}
	\IEEEauthorblockA{Univ Rennes, CNRS, IRISA\\ddealmei.0@gmail.com}
	\and
	\IEEEauthorblockN{Natalia Kulatova}
	\IEEEauthorblockA{Mozilla\\nkulatova@mozilla.com}
	\and
	\IEEEauthorblockN{Mohamed Sabt}
	\IEEEauthorblockA{Univ Rennes, CNRS, IRISA\\mohamed.sabt@irisa.fr}
	\and
	\IEEEauthorblockN{Pierre-Alain Fouque}
	\IEEEauthorblockA{Univ Rennes, CNRS, IRISA\\pa.fouque@gmail.com}
	\and
	\IEEEauthorblockN{Karthikeyan Bhargavan}
	\IEEEauthorblockA{INRIA Paris\\karthikeyan.bhargavan@inria.fr}
}

\maketitle

\input{sections/00_abstract.tex}

% \keywords{timing attacks; WPA3; Dragonfly; constant-time; hacl*}

\section{Introduction} \label{sec:introduction}
\input{sections/01_introduction.tex}

\section{Background} \label{sec:background}
\input{sections/02_background.tex}

\section{Leakage Exploitation} \label{sec:dictionary_attack}
\input{sections/03_leakage_exploitation.tex}

\section{Dragondoom: Side-channels in the cryptographic libraries} \label{sec:attack}
\input{sections/04_vulnerability.tex}

\section{Experimental Results} \label{sec:experimentation}
\input{sections/05_experimentation.tex}

\section{Mitigation and Disclosure}\label{sec:mitigations}
\input{sections/06_mitigations.tex}
\section{Discussion}\label{sec:discussion}
\input{sections/07_discussion.tex}

\section{Related Work}\label{sec:related_work}
\input{sections/08_related_work.tex}

\section{Conclusion} \label{sec:conclusion}
\input{sections/09_conclusion.tex}

\section*{Acknowledgments}\label{sec:acknowledgments}
\input{sections/acknowledgment.tex}

\bibliographystyle{IEEEtranS}
\bibliography{bibliography}

\appendices
\section{Attack on SAE-PT}
\label{app:vuln_saept}
\input{sections/appendix_vuln_saept.tex}

\section{Experiments on bin2bn}
\label{app:results_bin2bn}
\input{sections/appendix_res_bin2bn.tex}

\section{Experiments on set\_compressed\_point}
\label{app:additional_exp}
\input{sections/appendix_additional_exp.tex}

%%%%%%%%%%%%%%%%%%%%%%%%%%%%%%%%%%%%%%%%%%%%%%%%%%%%%%%%%%%%%%%%%%%%%%%%%%%%%%%%
\end{document}

%% file: resources/lstfstar.tex
\lstdefinelanguage{fstar}{%
  morekeywords=[1]{type,and,val,fun,let,in,ref,try,if,then,else,match,with,open,as,module,rec,end,assume,private,when,forall,ghost,assert,logic,array,pattern,effect,requires,ensures,decreases,modifies, abstract},
  morekeywords=[2]{WP, Post, Pre, PURE, DIV, STATE, EXN, ALL, TotST, M, GHOST, BAD, GTot, Lemma},
  morekeywords=[3]{},
  morekeywords=[4]{},
  morestring=[b]",
  sensitive=true,%
  numbersep=4pt,
  columns=[l]fullflexible,
  texcl=true,
  mathescape=true,
  identifierstyle={\sffamily},
  keywordstyle=[1]{\sffamily\maybecolor{dkblue}},
  keywordstyle=[2]{\sffamily\maybecolor{dkblue}},
  keywordstyle=[3]{\maybecolor{dkred}},
  keywordstyle=[4]{\rmfamily\itshape},
  rangeprefix=(*---\ ,
  includerangemarker=false,
  stringstyle=\ttfamily,
  showspaces=false,
  morecomment=[n]{(*}{*)},
  commentstyle={\itshape\maybecolor{dkred}},
  literate={->}{$\rightarrow\,$}{1}
           {<-}{$\leftarrow\,$}{1}
           {<>}{$\neq\,$}{1}
	   {delta}{$\delta$}{1}
	   {exists}{$\exists$}{1}
	   {forall}{$\forall$}{1}
      	   {True}{$\top$}{1}
      	   {False}{$\perp$}{1}           
           {\\in}{$\in\,$}{1}
           {~}{$\sim$}{1}           
           {'a}{$\alpha$}{1}
           {'b}{$\beta$}{1}
           {'c}{$\gamma$}{1}
           {tau}{$\tau\,$}{1}
           {STATE0}{$\mathsf{ST}^\prime\,$}{1}
	   {/\\}{$\wedge\;$}{1}
	   {\\/}{$\vee\;$}{1}
           {>=}{$\geq\ $}{2}
           {<=}{$\leq\ $}{2}
	   {<==>}{$\Longleftrightarrow \ $}{3}
	   {==>}{$\Longrightarrow \ $}{3}
           {_}{\textunderscore}{1}
           {--}{-}{1}   % Added due to         - not displaying properly.  Weird. ... but may make arrows look bad  {...}{$\ldots$}{3}
           ,
  breaklines=false}

%% file: sections/00_abstract.tex
%!TEX root = ../2022_dragondoom_to_dragonstar.tex

\begin{abstract}

It is universally acknowledged that Wi-Fi communications are important to secure. Thus, the Wi-Fi Alliance published WPA3 in 2018 with a distinctive security feature: it leverages a Password-Authenticated Key Exchange (PAKE) protocol to protect users' passwords from offline dictionary attacks. Unfortunately, soon after its release, several attacks were reported against its implementations, in response to which the protocol was updated in a best-effort manner.

In this paper, we show that the proposed mitigations are not enough, especially for a complex protocol to implement even for savvy developers. Indeed, we present \emph{Dragondoom}, a collection of side-channel vulnerabilities of varying strength allowing attackers to recover users' passwords in widely deployed Wi-Fi daemons, such as hostap in its default settings. Our findings target both password conversion methods, namely the default probabilistic hunting-and-pecking and its newly standardized deterministic alternative based on SSWU. We successfully exploit our leakage in practice through microarchitectural mechanisms, and overcome the limited spatial resolution of Flush+Reload. Our attacks outperform previous works in terms of required measurements.

Then, driven by the need to end the spiral of patch-and-hack in Dragonfly implementations, we propose \emph{Dragonstar}, an implementation of Dragonfly leveraging a formally verified implementation of the underlying mathematical operations, thereby removing all the related leakage vector. Our implementation relies on HACL*, a formally verified crypto library guaranteeing secret-independence. We design Dragonstar, so that its integration within hostap requires minimal modifications to the existing project. Our experiments show that the performance of HACL*-based hostap is comparable to OpenSSL-based, implying that Dragonstar is both efficient and proved to be leakage-free.

\end{abstract}

%% file: sections/01_introduction.tex
%!TEX root = ../2022_dragondoom_to_dragonstar.tex

\subsection{Context and Motivations}

Nowadays, there are more active Wi-Fi devices around the world than there are human beings. The Wi-Fi Alliance estimated the Wi-Fi global economy value to be \$3.3 trillion in 2021, and forecast its growth to \$4.9 trillion by 2025~\cite{wifi_economy21}. Such ubiquity and economic worth make protecting Wi-Fi vital. Since 2003, the Wi-Fi Alliance has introduced three major versions of its security protocol, namely the Wi-Fi Protected Access (WPA). The most recent one dates back to 2018, when WPA3 was announced in response to several serious weaknesses being identified in WPA2~\cite{DBLP:conf/uss/VanhoefP16,DBLP:conf/ccs/VanhoefP17,DBLP:conf/ccs/VanhoefP18}. A distinctive security feature of WPA3 is to leverage a Password Authenticated Key Exchange (PAKE) protocol, called Dragonfly, to protect users' passwords from the offline dictionary attacks that haunted the WPA2 handshake authentication. Despite its recent release, WPA3 has already been largely adopted by major Wi-Fi providers and software; in particular, it has become mandatory for Wi-Fi certification since July~2020~\cite{wifi_certification20}.

The rising popularity of Dragonfly, or its WPA3 variant called Simultaneous Authentication of Equals (SAE), and the controversy raised by several CFRG members~\cite{IETF:mail_archive/tls/M9Wrwd0iDEAk-PztgmrqIPEXvao,IETF:mail_archive/cfrg/WXyM6pHDjGRZXZzSc_HlERnp0Iw}, motivated research works to assess the security of its deployed implementations. Notably, Vanhoef and Ronen~\cite{DBLP:conf/sp/VanhoefR20} successfully found and exploited a microarchitectural side-channel in multiple WPA3 implementations to recover users' passwords. This attack was improved and extended to more implementations by De Almeida Braga \textit{et al.}~\cite{DBLP:conf/acsac/BragaFS20}. Both works target some secret-dependent execution within the controversial password derivation function of Dragonfly called \textit{hunting-and-pecking}~\cite{rfc7664}. Given their significance, multiple actions were taken following the disclosure of these attacks. Several countermeasures were applied to patch the vulnerable WPA3 implementations. In addition, the WPA3 standard has evolved to propose \textit{SSWU} (Simplified Shallue-Woestijne-Ulas)~\cite{DBLP:conf/crypto/BrierCIMRT10,irtf-cfrg-hash-to-curve-14}; an alternative password derivation function that is easier to securely implement.

After much effort, one might reasonably presume that the attack vectors as presented in~\cite{DBLP:conf/sp/VanhoefR20,DBLP:conf/acsac/BragaFS20} be no longer relevant. Namely, eminent open-source WPA3 implementations (\eg hostap) are expected to be exempt from microarchitectural side-channel attacks, especially that they were analyzed both manually and by leveraging automatic tools to detect microarchitectural leaks, such as MicroWalk~\cite{DBLP:conf/acsac/WichelmannMES18}. In this paper, we challenge this belief and raise the question again of whether Dragonfly is secure in practice. Our approach consists of analyzing and tracking password-tainted values within the password conversion operation of Dragonfly, especially inside calls to external functions. Indeed, Dragonfly implementations rely mostly on third-party libraries to perform their required cryptographic operations. These libraries do not always provide ``constant-time'' (secret-independent) implementations for their functions. Unfortunately, the security impacts of calling an external leaky function within Dragonfly was left unstudied. Our work brings to light this untapped source of leakage that can be successfully exploited to recover users' passwords even when SSWU is used.

\subsection{Our Contributions}
In this paper, we introduce \textbf{Dragondoom}: a collection of side-channel vulnerabilities caused by the supported cryptographic libraries in widely deployed  WPA3 implementations. Indeed, we found that most libraries we assessed do not execute Dragonfly operations independently from password-related values. For instance, the execution flow of the elliptic curve point decompression algorithm depends on the compression format, which is linked to the password, resulting in a side-channel leak. Unlike previous works, we show that the identified vulnerabilities do not only concern ``hunting-and-pecking'', but also the recently implemented SSWU that is secret-independent by design.

%Indeed, we show that widely deployed implementations still suffer from side-channel vulnerabilities caused by their supported cryptographic libraries. Indeed, Dragonfly implementations rely mostly on third-party libraries to perform their required cryptographic operations. These libraries do not always provide ``constant-time`` (secret-independent) implementations for their functions. Unfortunately, the security impacts of calling an external leaky function within Dragonfly was left unstudied. In our work, we analyze and track password-tainted values within the password conversion operation of Dragonfly, especially inside calls to external functions. Our analyses were rewarding; we notice that most libraries we assessed do not execute Dragonfly operations independently from password-related values. 
%For instance, the execution flow of the elliptic curve point decompression algorithm depends on the compression format, which depends on the used password, resulting in a side-channel leak. Moreover, 
%Unlike previous works, we show that vulnerabilities do not only concern the legacy ``hunting-and-pecking'', but also the recently implemented SSWU that is secret-independent by design.

We demonstrate our vulnerabilities in multiple cryptographic libraries used by widespread SAE implementations: hostap, iwd and FreeRadius. We describe how the leakage can lead to an offline dictionary attack, which defeats the purpose of leveraging a PAKE protocol in WPA3. To this end, we make use of well-known microarchitectural techniques, namely \fr and Performance Degradation Attack (PDA), to monitor some targeted instructions. We show the effectiveness of our attack by successfully recovering users' passwords in the most recent version of hostap using OpenSSL, which is the default setting for the Wi-Fi daemon on most Linux-based systems. We also demonstrate that other rising implementations, such as iwd/ell, are also affected by this vulnerability, to a greater degree.

%Due to the popularity of the targeted projects, we expect our vulnerability to impact millions, if not billions, of devices. We have not only communicated our findings to the maintainers of these open-source projects, but also reviewed their short-term patch. Note that the technical details of our work are the same in both WPA3 and EAP-pwd, since SAE is part of both protocols with only minor differences. For the sake of clarity, the core of the paper focuses on Wi-Fi daemons, since they are more prevalent.

Dragondoom highlights an interesting fact: implementation security in a well-established standard still does not withstand a simple side-channel analysis despite their prevalence. Therefore, rather than finding and fixing bugs in an ad-hoc manner, we propose \textbf{Dragonstar} as a long-term mitigation to prove the absence of large classes of vulnerabilities by design. Simply put, we consider hostap (one of the most deployed Wi-Fi daemons), and study its common cryptographic API that is used to process all the required low-level cryptographic operations. Then, we implement this API by leveraging the formally verified HACL* cryptographic library. Our design presents two main advantages. First, Dragonstar is easily deployable in practice since it relies on the flexible design of hostap. Second, it provides decent performance when compared to hostap calling legacy libraries, such as OpenSSL.

Due to the popularity of the targeted projects, we expect both Dragondoom and Dragonstar to impact millions, if not billions, of devices. We have not only communicated our findings to the maintainers of these open-source projects, but also reviewed their short-term patches. Note that the technical details of our work are the same in both WPA3 and EAP-pwd, since SAE is part of both protocols with only minor differences. For the sake of clarity, the core of the paper focuses on Wi-Fi daemons (\textit{i.e.}, SAE), since they are more prevalent. To summarize, our contributions are the following:

\begin{compactenum}
    \item We introduce \emph{Dragondoom}; a wide leakage vector caused by the supported cryptographic libraries.
    \item We examine the impact of Dragondoom in multiple projects, and show how it can result in passwords recovery. Here, we do not only target the legacy mode ``hunting-and-pecking'' like the previous work, but we also introduce the first side-channel vulnerability impacting the new SAE-PT mode.
    \item We provide a full Proof of Concept of our attack\footnote{\url{https://gitlab.inria.fr/ddealmei/artifact_dragondoom}} on the default Wi-Fi setting of most Linux systems (\textit{i.e.}, hostap with OpenSSL).
    \item We present \emph{Dragonstar}, an implementation of Dragonfly that uses formally verified code for all cryptographic operations,
      and we demonstrate how it can be used as drop-in replacement of OpenSSL in hostap without impacting performance\footnote{\url{https://github.com/ddealmei/dragonstar}}.
\end{compactenum}

In this paper, we show, once again, that implementing cryptographic protocols for real-world applications is error-prone, and can have both obvious and subtle vulnerabilities that are hard to detect. In particular, quite often, they rely on some third-party libraries to implement low-level cryptographic operations. The execution flow of these operations is not always independent of their arguments. Thus, a leakage occurs whenever a call with secret-dependent values is made. The diversity of protocols makes it unreasonable to strictly implement all these operations in constant time.

\subsection{Attack Scenario}
To illustrate the practical aspect of Dragondoom, we define the following potential attack scenario. We emphasize that we only describe a broad outline of \emph{one} potential scenario, and do not aim at giving all relevant details. Our focus is to prove that WPA3 is still vulnerable to side-channel attacks.

We assume Dragonfly is being used in WPA3, with a client (the victim) connecting to an Access Point (AP). The client would then use the Wi-Fi password to connect, whether for the first time (manually entering it), or automatically connecting to a known network. The goal of attackers is to recover the secret password. Similar to previous work, we suppose attackers controlling an unprivileged spyware running on the victim's device. The attack performs better if the victim is tricked to execute the WPA3 handshake with a fake AP controlled by attackers, so that more data can be collected on the same password. When enough data is leaked, attackers finally go offline to perform some dictionary partitioning attack. The adopted strategy is out of scope of our work.

\subsection*{Disclosure}
We disclosed our findings to the hostap security team in December 2021. We contacted other affected projects (iwd/ell from Intel and FreeRadius) in January 2022.
hostap promptly reacted, asking us to review a patch, which later was committed\footnote{Patch hostap~\href{https://w1.fi/cgit/hostap/commit/?id=6c380f4c87fbda0bc1a2c8d522412cf1fc01d477}{[1]}~\href{https://w1.fi/cgit/hostap/commit/?id=8c502336d40cdb8d211e4f8b8d59d7d282253733}{[2]}}, and a security advisory has been published. 
Intel decided to fix their cryptographic library, ell, and also asked us to review their patch\footnote{Patch ell \href{https://git.kernel.org/pub/scm/libs/ell/ell.git/commit/?id=f9a2d639ee349ef91a4ce34b41e4b6ac10997851}{[1]}~\href{https://git.kernel.org/pub/scm/libs/ell/ell.git/commit/?id=a9734ba128fe7353ef91796c893119266530c8fc}{[2]}}. Both iwd and hostap released a new stable version patching the vulnerability soon after our disclosure.
FreeRadius has committed our patch to their project.
We contacted OpenSSL and WolfSSL in May 2022 to disclose our second vulnerability. Both acknowledged our analysis, but argued that it is upon developers' responsibility to avoid calling their leaky functions with secret-dependent values.

%% file: sections/02_background.tex
%!TEX root = ../2022_dragondoom_to_dragonstar.tex

This section first gives an overview of the Dragonfly protocol and its variant Simultaneous Authentication of Equals (SAE) that is used in WPA3 and EAP-pwd. Then, we describe the side-channel techniques we leverage. Finally, we introduce the relevant notions of formal verification of secret independence regarding our work.

\subsection{The Dragonfly Key Exchange} \label{sub:dragonfly}

Dragonfly is a Password Authenticated Key Exchange (PAKE) protocol, designed by Dan Harkins in 2008~\cite{4622764}. It is an interactive protocol enabling two parties to establish an authenticated channel based on a supposedly low entropy secret. In particular, as a PAKE, it is required to resist (offline) dictionary attacks: the validity of a password can be guessed only by running a session and observing the outcome. Each part knows the shared secret before initiating the protocol, making Dragonfly a \emph{symmetric} PAKE. In particular, the protocol follows the same workflow for both sides.

After stirring some controversy during the CFRG review process~\cite{IETF:mail_archive/tls/M9Wrwd0iDEAk-PztgmrqIPEXvao,IETF:mail_archive/cfrg/WXyM6pHDjGRZXZzSc_HlERnp0Iw}, Dragonfly was properly described as RFC~7664~\cite{rfc7664} in 2015. Since then, Dragonfly and its variant, SAE, have been officially endorsed and widely deployed as key features by IEEE 802.11 and WPA3~\cite{9363693}, and used for EAP-pwd~\cite{rfc5931} and TLS-pwd~\cite{rfc8492}. %in informational RFC 

The handshake security relies on the discrete logarithm problem. Support for both multiplicative groups modulo a prime (MODP) and Elliptic Curve Cryptography over a prime field (ECP) is described in the references~\cite{rfc7664,9363693}. The exact operations of the handshake vary slightly depending on the underlying group. For the sake of brevity, we only consider ECP group. As required by the specifications (section 12.4.4.1 of~\cite{9363693}), we assume group 19 (corresponding to P256) as the default supported curve unless stated otherwise.

Thereafter, we adopt a classic elliptic curve notation: $G$ is the generator of a group, with order $q$. Lowercase denotes scalars and uppercase denotes group elements. For elliptic curve, we assume the equation to be in the short Weierstrass form $y^2 = x^3 + ax + b \bmod p$, where $a$, $b$ and $p$ are curve-dependent and $p$ is prime.

Dragonfly is broken into three parts: (i) password derivation; (ii) commitment; and (iii) confirmation. 

As the core of our contributions focuses on the password derivation, we only detail this step of the protocol.
The goal is, for both the sender and the receiver, to convert the shared password into a group element. Since IEEE 802.11-2020~\cite{9363693}, two methods are part of the standard. 

\paragraph{Hunting-and-pecking} The default method, presented as the only available option in previous versions of the standard~\cite{6178212,7786995}, is based on a probabilistic try-and-increment method called \emph{hunting-and-pecking}. 
This approach consists in hashing the password along with the identity of both parties and a counter until the computed value corresponds to a group element.
For ECP groups, this involves two steps. It first converts the password into the x-coordinate of a point. Then, since two y-coordinates may be valid, only one is chosen based on the parity of some values. The pseudocode describing this process on ECP groups is summed-up in \autoref{lst:h2c}.

\begin{lstlisting}[
	caption={\emph{Hunting-and-pecking} on ECP group as used in WPA3. The value of \code{label\_1} and \code{k} may vary along with the implementation, with a recommended value of \code{k}$\geq 40$.},
	label={lst:h2c},
	style=codesample,
	float,
	floatplacement=H,
	mathescape=true,
	escapeinside={@}{@},
	numbers=left,xleftmargin=2em,framexleftmargin=1.5em]
def hunting_and_pecking(pwd, macA, macB, ec):
	found, counter = False, 0
	A, B = max(macA, macB), min(macA, macB)
	while counter < k or not found:	@\label{lst:h2c_loopstart}@
		counter += 1
		seed = Hash(A || B || pwd || counter)
		x_cand = KDF(seed, label_$1$, ec.p) @\label{lst:h2c_kdf}@
		if is_x_coordinate(x_cand, ec) and not found:  @\label{lst:h2c_checkx}@
			x, pointType, found = x_cand, seed, True
			# Not described in the RFC, but implemented in SAE
			pwd = get_random(32) @\label{lst:h2c_loopend}@

	P = set_compressed_coordinate(x, 2 + (pointType & 1), ec) @\label{lst:h2c_sety}@

	return P
\end{lstlisting}

\paragraph{Hash-to-element} Following the disclosure of \textit{Dragonblood}~\cite{DBLP:conf/sp/VanhoefR20} in 2019, both the Wi-Fi Alliance and EAP-pwd updated their standard to describe a password conversion method based on SSWU~\cite{9363693,draft_harkins_eap_pwd_prime_00}. This method is a deterministic, efficient and easy-to-implement alternative to hunting-and-pecking. Readers can refer the pseudocode in \autoref{lst:h2e} and to the specification for more details.

\begin{lstlisting}[
	caption={\emph{Hash-to-element} on ECP group as used in WPA3. Capitalized variables denote points on the curve. The HKDF is based on the extract-then-expand paradigm. The first stage takes the input keying material and "extracts" a fixed-length seed.  The second stage "expands" the seed into several pseudorandom values of chosen length. Note that PT is computed once, and then reused at session establishment to compute P.},
	label={lst:h2e},
	style=codesample,
	float,
	floatplacement=H,
	mathescape=true,
	escapeinside={@}{@},
	numbers=left,xleftmargin=2em,framexleftmargin=1.5em]
def hash2element(pwd, ssid, identifier, ec):
	len =  olen(ec.p) + @$\lfloor$@ olen(ec.p)/2 @$\rfloor$@ 
	# extract seed material to get multiple random values
	seed = hkdf_extract(ssid, pwd || identifier)
	# expand the seed into a first random value
	pwd_value = hkdf_expand(seed, label_u1_P1, len) @\label{lst:h2e_u1}@
	u1 = pwd_value @$\bmod$@ ec.p 
	P1 = SSWU(ec, u1)	@\label{lst:h2e_P1}@
	# expand the seed into a second random value
	pwd_value = hkdf_expand(seed, label_u2_P2, len) @\label{lst:h2e_u2}@
	u2 = pwd_value @$\bmod$@ ec.p
	P2 = SSWU(ec, u2)	@\label{lst:h2e_P2}@
	PT = P1 + P2
	# PT is stored and used to generate a session specific P
	return PT
	
def get_password_element(PT, macA, macB, ec):
	k = hkdf_extract(0, macA || macB) 
	k = (k @$\bmod$@ (ec.q - 1)) + 1
	return k @$\times$@ PT
\end{lstlisting}

We stress that this second method is not backward-compatible. Therefore, both peers need to agree on the password derivation method to be used. The standard states that implementation should default to the original method, namely hunting-and-pecking, \textit{"If the AP does not indicate support for the SAE hash-to-element in its Extended RSN Capabilities field or the SAE initiator does not set the status code to SAE\_HASH\_TO\_ELEMENT in its SAE Commit message"}. 
Until recently, widespread implementations (such as hostap) did not implement it in their stable release. Upon disclosure of our vulnerability, hostap released a new version (v2.10), deploying a patch and support for SSWU-based hash-to-element.

\subsection{Simultaneous Authentication of Equals} 
\label{subsec:wpa3_deployement}

WPA3 includes a slight variant of Dragonfly, called SAE~\cite{9363693}. In this variant, the label values are fixed and each party is identified by its MAC address. Both peers are referred to as \emph{stations}, or \emph{STAs}. The protocol is referred as SAE or SAE-PT depending on the used password conversion method  (for hunting-and-pecking method and hash-to-element respectively).

The SAE handshake is executed between the client and the Access Point (AP) to compute the Pairwise Master Key (PMK). Afterward, a classic WPA2 4-way handshake is performed with this PMK to derive fresh cryptographic materials.
Since the entropy of this key is higher than in WPA2, the dictionary attack on the 4-way handshake is no longer relevant.

\subsection{Microarchitectural Preliminaries} 

\paragraph{Cache Architecture}
Caches are a special hardware feature found on modern CPUs to offset the discrepancy between fast processing and slow memory access. They act as small, but fast access buffers located directly on the CPU.
On modern hardware, caches are broken down into multiple levels, built with an access hierarchy, going from the lower L1 cache to the Last Level Cache (LLC), often shared across all cores. On personal computers, the LLC is usually L3. We note that L1 is the only cache where data and instructions are separated.% (into L1i and L1d respectively).

Each cache is divided into multiple sets, each of which contains multiple cache lines of fixed size (standard size is 64 bytes).
When executing a program, CPU cores will look into its lower cache for instructions and data. If the cache line is not available (cache \emph{miss}), it will be fetched from higher memory levels (L2, LLC, RAM). For later access, the CPU should find it in the cache (cache \emph{hit}), avoiding the overhead of accessing higher memory levels.

\paragraph{\fr Attack} \label{subsubsec:background-microarch-fr}
As CPU caches are shared among processes, attackers may leverage the time gap between a cache hit and a miss to get information on the cache state and infer a particular process access pattern. These attacks, referred as \emph{access-driven}, are particularly handy to guess the program execution flow. \fr has proven to be very accurate and efficient to monitor cache access using the state of the inclusive LLC (inclusive caches are commonly found on Intel processors).

The idea is to exploit memory sharing between two processes - a victim $\mathcal{V}$ and a spyware $\mathcal{S}$. The goal of $\mathcal{S}$ is to probe a specific memory line from this shared memory. The attack is carried out in three steps. First, $\mathcal{S}$ flushes the probe out of the LLC with the appropriate instruction (\code{clflush} or \code{clflush\_opt}). Since the LLC is inclusive, this operation ensures that the probe is no longer in any cache. $\mathcal{S}$ then idles for a period of time, during which it waits for $\mathcal{V}$ to execute - or not - the probe. Finally, $\mathcal{S}$ reloads the probe and measures the reload time. If $\mathcal{V}$ has executed the probe during the idle period, the reload results in a hit, hence a short reload time. Otherwise, the access will be significantly longer. Repeating these steps allows the attacker to identify when the probe has been used by the victim process.

\paragraph{Performance Degradation Attack (PDA)}
The mere idea behind \fr is to perform a few assembly instructions to flush, reload and time the operations. Naively looping over these operations would introduce blind spots on concurrent reload, where attackers would miss many events. To prevent this, measurements are usually performed periodically, by defining an appropriate idle period between flushing and reloading.
Doing so, on one hand, we increase the temporal resolution by the additional idle. 
On the other hand, attackers would not distinguish multiple accesses to the probe if they occur within the same time slot. The PDA is a technique in which a well-chosen memory line is constantly flushed out. This allows attackers to benefit from the longer idle without missing valuable information~\cite{DBLP:conf/acsac/AllanBFPY16}.
Combining \fr with PDA is common, and has been leveraged numerous times~\cite{DBLP:conf/acsac/AllanBFPY16,DBLP:conf/ccs/GarciaBY16,DBLP:conf/uss/GarciaB17,DBLP:conf/ches/BernsteinBGBHLV17,DBLP:conf/ccs/PesslBY17,DBLP:journals/tches/AldayaGTB19,DBLP:conf/ccs/AranhaN0TY20,DBLP:conf/sp/CohneyKPGHRY20,DBLP:conf/uss/GarciaHTGAB20,DBLP:conf/acsac/BragaFS20,DBLP:conf/ccs/BragaFS21}.

\subsection{Formally Enforcing Secret Independence}
\paragraph{Secret Independence}
One of the most common sources of side-channel leaks in cryptographic code is
secret-dependent control flow, \eg when an implementation branches on
a value derived from a secret, or performs secret-dependent memory access,
\eg when an array is indexed with a secret value. Such leaks have
been successfully exploited to mount remote timing attacks
on implementations of a variety of cryptographic constructions.

To prevent such attacks, cryptographic experts recommend a coding
discipline that is sometimes (somewhat misleadingly) termed
``constant-time'', and is more accurately called \emph{secret
independence}: all branches and memory accesses should
only be dependent on public values. This discipline is rigorously
implemented in mainstream cryptographic libraries, but is hard
to get right, as the attacks in this paper demonstrate.

\paragraph{Formal Verification of Secret Independence}
Several works have proposed semi-automated methods for formally
verifying that cryptographic code is secret independent.
See \cite{cac-sok} (Section IV) for a recent survey; in particular,
tools like Vale~\cite{vale2} and Jasmin~\cite{jasmin} can analyze
secret independence for assembly code, while HACL*~\cite{hacl-star}
enforces secret independence on cryptographic implementations in C.
Each method relies on a formal leakage model, where attackers are
typically allowed to observe all branch results and the sequence of
all memory addresses accessed by the program. A verification tool then
implements a sound and conservative analysis, so that if it says a
program is secret independent, then attackers cannot distinguish a
secret from a fresh random value.

\paragraph{HACL*: A Verified Cryptographic Library}
HACL* is a cryptographic library~\cite{hacl-star,evercrypt,haclxn}
that includes formally verified C implementations for a full suite of
modern cryptographic algorithms, including hash functions, encryption
algorithms, elliptic curves, and signature schemes, all of which are
verified for functional correctness, memory safety, and secret
independence. Each cryptographic implementation is written in the F* programming
language~\cite{fstar}, verified using the dependent type system of F*,
and then compiled to C via a tool called KreMLin~\cite{kremlin}.  A
compilation theorem guarantees that all verified guarantees, including
secret independence, hold in the generated C code~\cite{kremlin}.

When implementing a cryptographic algorithm in HACL*, the programmer
annotates each bytestring and integer in the program as secret or
public.  This F* type system tracks all secret and public bytes and
integers to enforce a secret-independent discipline. For example,
each encryption key is treated as an array of secret (opaque) bytes.  The
programmer may convert these bytes into secret integers, perform
secret-independent operations like additions and multiplications,
and then convert them back to bytes.  However, the programmer
cannot compare secret bytes (or integers) and cannot use secret
integers as array indexes. A security theorem then states that this
type-based discipline enforces secret independence~\cite{kremlin}.

%% file: sections/03_leakage_exploitation.tex
%!TEX root = ../2022_dragondoom_to_dragonstar.tex

\subsection{Previous Attacks}
Side-channel attacks against WPA3 can be divided into four steps: (1) finding some password-dependent execution, (2) identifying the leakage source of password-related data, (3) collecting data through side-channels, and (4) performing an offline dictionary partitioning attack. Related attacks mostly share step 4, and differ in other steps. The strategy of recovering passwords, namely dictionary partitioning, is often considered as a possible implication of the identified leakage, and not a purpose in its own. Indeed, step 4 aims at providing a common benchmark about the efficiency of the identified leakage. We do not claim that this approach is the best way to recover passwords. Indeed, dictionary attacks do not necessarily lead to password recovery, especially for strong passwords. However, we still follow this approach to build a common benchmark with previous attacks. As for step 3, it depends on the leveraged microarchitectural techniques, while step 2 defines the number of bits related to the secret per execution trace. Previous attacks~\cite{DBLP:conf/sp/VanhoefR20,DBLP:conf/acsac/BragaFS20} share steps 1 and 4, but differ in steps 2 and 3. Indeed, both exploits the password-conversion loop (lines~\ref{lst:h2c_loopstart} to \ref{lst:h2c_loopend} of \autoref{lst:h2c}). Below, we describe their differences.

The Dragonblood attacks~\cite{DBLP:conf/sp/VanhoefR20} presented a side-channel on the password conversion. In steps 2 and 3, they recover information on the password by observing the duration of the first iteration of the password conversion; a successful conversion would need more time to process the additional instructions. A successful conversion occurs with a probability of 0.5. Hence, this attack provides a fixed amount of leakage on the used password.

In~\cite{DBLP:conf/acsac/BragaFS20}, authors perform a similar attack on implementations that did not properly mitigate Dragonblood. They improved the original attack in step 2 by targeting more specific instructions to extract more information from each observation. Namely, instead of only learning whether the successful conversion occurs during the first iteration, they get the exact corresponding iteration. Considering the conversion probability at each iteration, this doubles the amount of obtained leakage.

Our attack differs from these works in multiple aspects. First, in step 1, we identify a \emph{new leakage source}. Previous attacks leak information about the conversion iteration, and they were patched by making the workflow of each iteration of the loop secret-independent. Our attack instead targets generic functions from the called cryptographic library (line~\ref{lst:h2c_sety} of \autoref{lst:h2c}). To the best of our knowledge, this leakage vector has not been identified before. 
Second, experimentally, our attack in step 3 requires fewer side-channel measurements than previous works to recover a password for the same threat model. Being implementation-dependent, we will provide more details about the attack practical efficiency in \autoref{ssec:compare_experimental}. We achieve this performance by introducing a different Flush+Reload-gadget (see \autoref{ssubsec:fr_gadget}), leveraging instruction prefetching to overcome the spatial limitation of classical attacks.
More importantly, unlike previous works, Dragondoom provides a wider \emph{scope} for attackers, impacting not only all previously patched implementations, but also the recently supported SAE-PT. We show that despite its constant-time design, it still suffers from implementation pitfalls. In follows, we present steps 3 and 4 in a generic way, as well as the threat model.

\subsection{Threat Model}
\label{subsec:threat_model}
As is often the case when targeting Wi-Fi daemons, we consider attackers with some physical proximity (\textit{i.e.}, within network range). We also assume that they can monitor multiple handshakes using the same password while varying at least one MAC address. This can be achieved differently depending on the target. First, suppose the target be an AP. In that case, attackers can try to connect, triggering an (invalid) handshake (the relevant part of the handshake will execute even if the password is invalid). Second, if the target is a client, attackers can set up a fake AP to impersonate a valid one (known to the user) by spoofing the original AP and advertising the same SSID with stronger signal strength (making it the default choice for the client Wi-Fi daemon). In both cases, attackers can easily control the MAC address used by the targeted device in order to leak enough data to recover the password. Setting a fake AP comes with other benefits. Indeed, they can force authentication using SAE, with the default password derivation, by simply omitting the corresponding Extended RSN Capabilities field. Moreover, we also assume that attackers control an unprivileged process running on the victim's device. We do not consider an attacker having access to the victim's Wi-Fi. Therefore, our model applies even when the spy process is sandboxed, or access to the network manager (\eg nmcli on Linux and netsh on Windows) has been restricted by some access control rules. For instance, by default on some operating systems, processes running on behalf of an unprivileged user cannot read the Wi-Fi passwords entered by another user. Our attack still works in these restricted cases because the only requirement is to be able to execute some unprivileged instructions. Finally, we stress that our threat model is common to previous cache-based attacks~\cite{DBLP:conf/acsac/BragaFS20,DBLP:conf/sp/VanhoefR20}.

\subsection{Step 3: Collecting Data and Leakage Aggregation}
\label{subsec:fingerprint}
Side-channel leakage samples are obtained through observing some execution. We recall that the two standardized password conversion methods are deterministic with respect to the password and the sender's/receiver's identities. This means that, except for masked computations, we can observe the same values for the same run. This is fortunate for the attackers, since microarchitectural techniques are noisy, thereby requiring several observations to obtain reliable leakage.

Once the leakage is acquired, attackers might need more data than one run can reveal to succeed in practice. Therefore, they may trigger new leakage on the same secret value. This is possible only when the leaked secret-related value involves public parameters that can be maliciously modified.

Thus, given a spying process leaking $\ell$ bits of a secret value, we denote $\ell_i$ as the leakage related to a given execution, namely a fixed set of public parameters. Consequently, the overall leakage obtained by attackers can be represented as follows: $\mathcal{C}_n = \left\{\ell_i, i = 1...n \right\}$. We say that $\mathcal{C}_n$ contains $n$ traces. Ultimately, $\mathcal{C}_n$ may be used as a fingerprint of the password to perform a dictionary partitioning attack.

\subsection{Step 4: Dictionary Partitioning Attack} 
\label{subsec:dictionary_attack}

Once the fingerprint of the targeted password is leaked, attackers can switch into an offline mode. Indeed, they apply the password conversion function to a dictionary of password candidates. Then, they discard each candidate whenever the leaked bits do not match the computed values. In the case of hunting-and-pecking, they can drop all dummy computations once the password is converted.

An interesting question is how large the fingerprint should be to uniquely find the correct password. In other words, supposing a dictionary with $d$ passwords, what is the smallest number of traces that eliminates all wrong candidates with high probability? Intuitively, given a fingerprint $\mathcal{C}_n$, the probability that a password candidate does not match $k$ leakages follows a binomial distribution, with $p$ being the probability corresponding to the leakage. Thus, we can infer the probability that $d$ candidates are pruned given $n$ traces. The smaller $n$, the more efficient the attack is. For a complete description, we defer the reader to~\cite{DBLP:conf/sp/VanhoefR20} (Section 7.2) and~\cite{DBLP:conf/acsac/BragaFS20} (Section 3.5).

Simply put, the fingerprint $\mathcal{C}_n$ allows attackers to perform an offline dictionary attacks, as demonstrated in previous attacks against WPA3~\cite{DBLP:conf/sp/VanhoefR20, DBLP:conf/acsac/BragaFS20}. In our paper, we propose to extend this benchmark measuring the efficiency of the resulted offline dictionary attack, while not involving the dictionary size. Thus, we introduce $\mathcal{D}$ as the ratio between the leakage for a fixed set of parameters ($\ell$), and the number of measurements necessary for a reliable leakage ($\mathcal{R}$). This is directly related to the number of required microarchitectural measurements to prune a given dictionary. A perfect dictionary attack would allow attackers to prune all invalid candidates by observing one single online authentication session. This would translate to $\mathcal{R} = 1$ and $\ell=\log_2(\#dictionary)$, hence $\mathcal{D} = \log_2(\#dictionary)$. Generally speaking, the bigger $\mathcal{D}$ is, the better the attack is. Our experimental results, detailed in \autoref{sec:experimentation}, show that our identified vulnerabilities against hunting-and-pecking outperforms previous attacks defined in~\cite{DBLP:conf/sp/VanhoefR20} and~\cite{DBLP:conf/acsac/BragaFS20}.

%% file: sections/04_vulnerability.tex
%!TEX root = ../2022_dragondoom_to_dragonstar.tex

In this section, we show how Dragonfly implementations can still leak during the password conversion method. In particular, we investigate calls to external cryptographic libraries with password-related values. We highlight two sources of leakage, which we call \emph{Dragondoom}, affecting both the default hunting-and-pecking and the recently standardized SSWU-based hash-to-element. The identified leakages are related to passwords, so they can directly result in dictionary attacks as discussed in \autoref{sec:dictionary_attack}. This section describes the theoretical study of these interactions (steps 1 and 2 of the attack), while practical aspects are discussed in \autoref{sec:experimentation}.

\subsection{Motivation}
Along the standardization process, several design flaws have been identified against widely deployed WPA3 implementations because of hunting-and-pecking~\cite{IETF:mail_archive/tls/M9Wrwd0iDEAk-PztgmrqIPEXvao,DBLP:conf/sp/VanhoefR20,DBLP:conf/acsac/BragaFS20}. Consequently, various mitigations have been implemented to avoid password-dependent time variation in the execution of this password conversion method. The mitigations include: (1) fixing the number of iterations to 40, (2) performing any additional operation on a dummy random string instead of the password, (3) implementing the loop in constant time, and (4) masking the Legendre symbol computation. All these modifications show that it is hard to provide secure implementations of hunting-and-pecking, even for savvy developers.

Despite much effort, we suspect that WPA3 implementations still leak passwords. Our intuition comes from the fact that previous works only discovered leakages caused by the WPA3 code itself; in particular the password-conversion loop (lines~\ref{lst:h2c_loopstart} to \ref{lst:h2c_loopend} of \autoref{lst:h2c}). However, most WPA3 implementations rely on external libraries to perform some mathematical operations on big numbers or elliptic curves. Some of the called functions are not secret-independent, because they mostly deal with public values. Nevertheless, they do involve password-dependent values in Dragonfly.

In this paper, we look carefully into the part of the code making external calls with secret-related values. Here, we perform a tainted analysis of the password conversion function using timecop\footnote{\url{https://post-apocalyptic-crypto.org/timecop/}}, combined with manual analysis. We highlight two main leakages. First, the point decompression, called when the hunting-and-pecking is successful, leaks a relation between a point coordinate and a password-related value. Second, the binary-to-big-number routine, called on various values including secrets, leaks the number of MSB of the input being zeros. Despite the previous extensive and systematic study of the big number operations~\cite{DBLP:conf/uss/WeiserSBS20}, these functions were not exploited before.

Of particular interest, we study widely used libraries, namely OpenSSL, WolfSSL and ell, supported by popular SAE (and EAP-pwd) implementations: hostap, FreeRadius and iwd. Our findings reveal that all of these libraries are affected by at least one vulnerability. Consequently, projects relying on the generic point decompression function exposed by these cryptographic libraries leak secret information. One peculiar example is Apple CoreCrypto which provides its own SAE implementation, where it re-implements routines to be secret-independent. A deeper analysis of all these implementations and their differences is given in \autoref{sec:experimentation}. Note that our work might be of interest for any protocol calling the analyzed leaky functions with secret-dependent arguments. In this paper, we focus solely on the resulting vulnerabilities in WPA3.

\subsection{Point Decompression}
\label{subsec:point_decompression}

This vulnerability only affects hunting-and-pecking (SAE), as the routine is not used in SSWU (SAE-PT).

\paragraph{Leakage Data: Seed Parity}
We recall that, after the conversion loop, we expect to have found the x-coordinate of the resulting point with overwhelming probability. Since a single x-coordinate can describe two points ($(x, y)$ and $(x, p-y)$), the algorithm requires a deterministic way to choose the output. Here, Dragonfly relies on the LSB of the seed value (\textit{i.e.}, its parity) that generated the x-coordinate, to determine the compression format of \code{set\_compressed\_point\_coordinate} (\autoref{lst:h2c}, line~\ref{lst:h2c_sety}).

Note that the seed is computed from one secret value and two public ones: the password and MAC addresses of each end. As a result, any bit of information leaked from the seed can be related to the password, thereby causing a dictionary partitioning attack (cf. \autoref{subsec:dictionary_attack}). Thus, the security of WPA3 also relies on whether the point decompression algorithm is secret-independent regarding the seed parity. Next, we describe the internals of this algorithm as applied for elliptic curves supported by SAE.

\paragraph{Leakage Origin: Compression Format}

\begin{lstlisting}[
	caption={Point Decompression Algorithm},
	label={lst:point_decompress},
	style=codesample,
	float,
	floatplacement=H,
	mathescape=true,
	escapeinside={@}{@},
	numbers=left,xleftmargin=2em,framexleftmargin=1.5em]
def set_compressed_point_coordinate(x, pointType, ec):
	y_sqr = x@$^3$@ + ec.a@$\times$@x + ec.b @$\bmod$ @ec.p
	y = sqrt_mod(y_sqr, ec.p)
	if ( y @$\bmod 2 \not\equiv$@ pointType @$\bmod 2$@ ) @\label{lst:point_decompress_if}@
		y = ec.p - y @\label{lst:point_decompress_sub}@
	P = (x, y) @\label{lst:point_decompress_setPoint}@
	
	return P
\end{lstlisting}

As described in \autoref{lst:point_decompress}, the generic decompression algorithm takes three parameters: the x-coordinate of a point, \code{pointType} as the compression format, and \code{ec} that stores the curve parameters. The algorithm computes one of the two candidates for y-coordinate, using Tonelli-Shanks. Then, it selects the y-coordinate to return based on the value of \code{pointType} and the parity of y. We note that because $p$ is an odd prime number, only one y candidate can be even. On a naive implementation, it is easy to notice that the branch line~\ref{lst:point_decompress_if} would leak whether the parity of $y$ matches the parity of the compression format. Interestingly, while this line expresses a single condition, it may be broken down into multiple branches. For instance, one can implement it by looking first at the compression format, and then processing it differently depending on the parity of $y$.

\paragraph{Leakage Impact}
The point decompression method is called with two password-tainted values: the x-coordinate and the compression format. Obviously, we suppose that attackers do not know either value and cannot make a knowledgeable guess about their parity. However, we suppose that attackers be able to guess whether conditional subtraction in line~\ref{lst:point_decompress_sub} of \autoref{lst:point_decompress} is executed, which implies that they would recover some information about the parity of secret values. The x-coordinate being uniformly distributed on the curve, the parity of y also is. Moreover, since the seed is the output of a cryptographic key-derivation function, it can be seen as the output of a random oracle. Hence, it is plausible to consider that their parity is equal with probability $pr = 0.5$.

This means that attackers can recover one bit of information if any leakage in line~\ref{lst:point_decompress_sub} occurs. A second bit of information is recovered if the condition in line~\ref{lst:point_decompress_if} is broken into multiple steps: the parity of $y$ or the seed, with the parity equality of both values. 

\subsection{Binary to Big Number Conversion}
\label{subsec:bin2bn}

This vulnerability affects both hunting-and-pecking (SAE) and SSWU-based hash-to-element (SAE-PT).

\paragraph{Leakage Data: Secrets MSB}
In cryptography, it is quite common to perform computations on integers not fitting in native types (usually limited to 64 bits on modern architectures). Thus, libraries define a special structure, called "big numbers" (abbreviated BN from now), that often boils down to a set of buffers representing the actual value, alongside flags and indexes. However, the manipulated values do not always present as BN; they can come in other formats. Thus, they require to be parsed before any computation. Of particular interest, coordinates of the secret point and secret-dependent values shall also be appropriately parsed, or converted, into the BN structure. Below, we note that the secret values can be leaked if the conversion routine is not secret-independent.

For SAE (\ie hunting-and-pecking), a coordinate candidate is computed at each iteration as the output of a KDF (\autoref{lst:h2c}, line~\ref{lst:h2c_kdf}), that is then converted to a BN. 
At each iteration, this value is generated like the seed, with additional deterministic processing. Hence, the same consequences follow (c.f. \autoref{subsec:point_decompression}).

For SAE-PT (\ie SSWU), implementations leak when the coordinates computed by SSWU are set (in the function call line~\ref{lst:h2e_P1} and \ref{lst:h2e_P2} of \autoref{lst:h2e}). The leak may also concern the input to these function calls, denoted \code{u1} and \code{u2} in the code sample, as it is the output of the HKDF (\autoref{lst:h2e} line~\ref{lst:h2e_u1} and line~\ref{lst:h2e_u2}). All leaked values are computed from the SSID, the password identifier and the secret password. 

\paragraph{Leakage Origin: Optimized Conversion}

\begin{lstlisting}[
	caption={Binary to Big Number Algorithm},
	label={lst:bin2bn},
	style=codesample,
	float,
	floatplacement=H,
	mathescape=true,
	escapeinside={@}{@},
	numbers=left,xleftmargin=2em,framexleftmargin=1.5em]
def bin2bn(buf, n):
	# Skip leading 0's
	while (buf[0] == 0) @\label{lst:bin2bn_opti}@
		n--
		buf++
	
	bn = new_bn()
	while(n--)
		add_byte_to_bn(buf, bn)
		
	return bn
\end{lstlisting}

As described in \autoref{lst:bin2bn}, we consider the conversion function to take two arguments: the binary buffer and its byte-length. Values in the buffer are considered secret. 
The routine is straightforward: after some sanity checks (not displayed in the code sample for conciseness), it converts the bytes in \code{buf} into chunks in \code{bn} and returns the result.

To avoid processing leading bytes to zero, which does not affect the final value, a quite common optimization is to skip them before proceeding to conversion, thereby decreasing the size of the buffer accordingly. Attackers able to guess the number of iterations in the loop line~\ref{lst:bin2bn_opti} can thus deduce the effective byte-length of the secret.

\paragraph{Leakage Impact}
The binary conversion function is called on several occasions during the password conversion process. The values we consider are uniformly distributed on the range $[0, 2^{n*8})$ (with negligible bias). Therefore, the $k$-leading bytes are zero with probability $pr = 1/256^k$, leaking $k\times8$ bits of information.

For SAE (\autoref{lst:h2c}), three values are underlined: (i) the first x-candidate; (ii) the final x-coordinate; and (iii) its corresponding y-coordinate. The MSB of (i) and (ii) may not be not independent, while MSB of (iii) is. Indeed, attackers can leak information on x-candidates as long as the observation they make is different from the one made on the final x-value. This allows attackers to determine whether the conversion was successful at the first iteration, in addition to the MSB leakage.

In more detail, we distinguish two cases depending on whether the leakage from (i) and (ii) is equal or distinct. As a reminder, each iteration may convert the password successfully with probability $q/2p\approx0.5$, with $p$ and $q$ being parameters of the curve.
In the case of equality, attackers have to consider these values as equivalent. Both values share a leading byte to zero if they are equal (successful conversion at the first iteration occurs with probability 0.5) or if they are different but share a common leading byte. The probability of such an event is $pr=0.5\cdot(1/256 + 1/256^2)$, in which case attackers would leak a total of 8 bits of information from both observations.
Otherwise, if the leaked bits differ, attackers can infer that both values are different. Thus, they can deduce that the password was not converted during the first iteration. This event occurs with probability 0.5, leaking an additional bit of information. In the end, attackers learn 9 bits of information with probability $pr=255/256^2$.
Similar reasoning can be applied to subsequent iterations, but the probability of leak quickly becomes negligible.

As for SAE-PT (\autoref{lst:h2e}), two points are created from the password. For each point, the two coordinates are converted to BN when the resulting point is set. Each coordinate is uniformly distributed on the curve, hence each call would leak 8 bits with probability $1/256$. The same observation applies to the input of SSWU, resulting in three independent leaking values at each call of SSWU. Since it is called twice per run, we can expect a $k\times8$-bit leakage with probability $pr=6/256^k$.

%% file: sections/05_experimentation.tex
%!TEX root = ../2022_dragondoom_to_dragonstar.tex

The attack condition exploited in \autoref{sec:attack} emerged from different implementation weaknesses within three popular projects: hostapd, FreeRadius and iwd. These projects mostly rely on three cryptographic libraries for their SAE (and EAP-pwd) operations: OpenSSL, WolfSSL and ell. In this section, we study the details of the different \code{set\_compressed\_point\_coordinate} and \code{bin2bn} routines. We find all three libraries to be affected by at least one of the vulnerabilities we introduced in \autoref{sec:attack}. While our identified flaws are exploitable in practice in all studied implementations, we only conduct our experiments in real-world settings against \code{wpa\_supplicant} interacting with OpenSSL, which is the default installation in most Linux distributions. Following the same approach, we also demonstrated the 2-bit leakage on ell and a 1-bit leakage on WolfSSL. For the sake of clarity, we detail our experiments for OpenSSL only, and defer the reader to \aref{app:additional_exp} for the additional experimental results. 

Moreover, since SAE-PT is not widely deployed yet, we solely focus on SAE with hunting-and-pecking in this section. We also introduce a practical attack scenario and perform a comparative study with previous works. We defer our presentation of the impact of our attack on SAE-PT to \aref{app:vuln_saept}.

\input{resources/pt_decompress_code.tex}

\subsection{Vulnerable Implementations}
As stated in \autoref{sec:attack}, the implementations of vulnerable routines may vary between libraries. Here, we analyze four open-source implementations of SAE, and notice that a third-party library usually provides the routines. We evaluated several open-source cryptographic libraries supported by the studied SAE projects. We did not include all the supported libraries. Instead, we only consider those for which elliptic curve operations are implemented since SAE specifically requires their support (with group 19 as a minimal requirement~\cite{9363693}). Moreover, we did not look into any proprietary project, such as the one running on Windows. The list of evaluated implementations is summarized in \autoref{table:sae_crypto_lib}. We also note the average number of leaked bits for each handshake execution.

\input{resources/table_affected_projects}

All routines for \code{set\_compressed\_point\_coordinate} and \code{bin2bn} of OpenSSL~\cite{openssl}, WolfSSL~\cite{wolfssl} and ell~\cite{ell} are described in \autoref{lst:practical_implem}. Below, we go through all these implementations to explore their flaws. We intentionally exclude Apple CoreCrypto~\cite{corecrypto} routines, although its generic point decompression function also leaks information on the compression format. CoreCrypto, unlike other libraries, provides its own implementation of SAE, where the decompression is re-implemented internally in a secret-independent fashion. In addition, the optimization causing the second vulnerability is not used. This prevents our attacks. We do not claim anything about the SAE running in Apple systems, since this requires reverse engineering efforts to determine whether it leverages CoreCrypto SAE. Next, we address the actual leakage in each implementation.

\noindent\textbf{OpenSSL (\autoref{lst:openssl_impl}).} The point decompression implements the naive approach described in \autoref{subsec:point_decompression}, which leaks one bit of information for each execution, namely whether the parity of $y$ is equal to the parity of \code{pt}. The condition line~\ref{lst:openssl_impl-y0} might also leak if $y$ is zero, which happens with negligible probability for a random point. 
The binary conversion skips the leading zero bytes (line~\ref{lst:openssl_impl_bin2bn-opti}). In hostap, this function is called on the three different secrets described in \autoref{subsec:bin2bn}, thereby leaking $8$ bits of information with probability $p=1/256 + 0.5\cdot(1/256 + 1/256^2)$, and 9 bits with probability $p=255/256^2$ in each session (avg. 0.094 bits per session). % event = [1/256: MSB(y) = 0, 255/25:6 MSB(y) != 0, 0.5*(1/256+1/65536): MSB(x_i) == MSB(x) == 0, 0.5*(255/256 + 255^2/256^2) MSB(x_i) == MSB(x) != 0, 255/65536: MSB(x_i) != MSB(x)]
In FreeRadius, the routine is called on an additional independent secret, leaking more data (avg. 0.131 bits per session).

\noindent\textbf{WolfSSL (\autoref{lst:wolfssl_impl}).} The point decompression leaks the same information as OpenSSL. However, the \code{if} statement (line~\ref{lst:wolfssl_impl-if1}-\ref{lst:wolfssl_impl-if2}) is valid if either condition is valid. Hence, upon execution of line~\ref{lst:wolfssl_impl-mod}, attackers might guess which condition was valid by observing if the function \code{mp\_isodd} was executed once (\textit{i.e.}, $y$ is odd) or twice (\textit{i.e.}, $y$ is even). Here, attackers would learn the parity of both $y$ and \code{pt}, leaking two independent bits of information.
The binary conversion leaks exactly the same amount of information as OpenSSL does, because of the optimization in line~\ref{lst:wolfssl_impl_bin2bn-opti}.\\

\noindent\textbf{ell (\autoref{lst:ell_impl}).} On one hand, the point decompression leaks the most, since it goes through all the point decompression in a switch over \code{pt}. Attackers can spy on the executed case to learn \code{pt} parity. Then, they can spy on the inner condition to guess the parity of $y$. Thus, they can learn two independent bits of information for each handshake execution.
The binary conversion, on the other hand, does not leak regarding the MSB of its input.

\subsection{Practical Attack Against wpa\_supplicant}
\label{ssec:practical_attack}

Now, we show the severity of the vulnerabilities by recovering users' passwords in a complete scenario with the default settings in most Linux distributions. As shown in \autoref{table:sae_crypto_lib}, the vulnerability caused by point decompression is more practical, as it leaks more bits. Hence, we detail our results regarding point decompression and defer BN conversion presentation to \aref{app:results_bin2bn}.

\subsubsection{Motivations and Setup}
\label{subsubsec:exp-attack-setup}
We illustrate the effectiveness of our attack by targeting wpa\_supplicant version 2.9, relying on OpenSSL version 1.1.1l, both being the most recent version at the time of testing. This choice was guided to define our experiment settings as close to default users' installations as possible. Indeed, OpenSSL is arguably the most deployed open-source cryptographic library, installed by default in many Linux distributions. Similarly, wpa\_supplicant is the default Wi-Fi daemon on Debian-based systems (Ubuntu, Kali, etc.), Android, and Red Hat systems for the authentication phase. Please refer to \aref{app:results_bin2bn} and \aref{app:additional_exp} for further experiments on \code{bin2bn} on OpenSSL as well as \code{set\_compressed\_point\_coordinate} on WolfSSL and ell.

All tests were performed on a Dell XPS 13 7390 running on Ubuntu 20.04.2, kernel 5.13.0-39, with an Intel(R) Core(TM) i7-10510U and 16 GB of RAM. All binaries were compiled with gcc version 9.4.0 with all default configurations (optimization included).

We considered the threat model described in~\autoref{subsec:threat_model}. Namely, we deployed an access point on an Android phone (running Android 11 on a kernel v4.9.227-perf+) to share network access. In addition, we kept the default configuration on both ends, meaning the key exchange uses group 19, corresponding to P256. Similar results would have been observed with other curves. 
Our spy process has been implemented using Mastik v.0.1-alpha~\cite{Mastik} to run the \fr and PDA processes.

\subsubsection{Flush+Reload Gadget}\label{ssubsec:fr_gadget}

We recall that attackers need to successfully determine whether line~\ref{lst:point_decompress_sub} of \autoref{lst:point_decompress} has been executed. 
Despite being easily identifiable in the source code, recovering this leak is challenging, since the difference in the control flow only represents a few instructions during the overall execution. Indeed, in the \code{set\_compressed\_point\_coordinate} function, the secret-dependent branch to spy on is quite small; consisting of a single call to an atomic arithmetic operation. 

\paragraph{Limitation of Flush+Reload}
The naive approach would be to probe the instructions inside the branch in order to detect their execution. Attackers would then periodically flush and reload the memory line corresponding to the shared instructions. A cache hit is observed only when the victim uses the probe. 
To make leakage more reliable, and measurements easier to acquire, it is common for attackers to increase the workload by performing PDA on some operations~\cite{DBLP:conf/acsac/AllanBFPY16}. Here, we notice that the probed line is just a few instructions after the branch. 
This implies that when the execution gets to the branch, the instructions inside the branch might be prefetched to the CPU execution pipeline (thus into the cache) to save some time in case the branch condition is satisfied. This prefetching may also be caused by the spatial locality of the memory lines, since it can be adjacent to the cache line computing the branch condition.
Hence, we expect, and empirically observe, that the probed instructions are present in the cache, even without being actually executed by the victim. This results in a high false-positive ratio. Similar observations and studies about spatial limitations of \fr were considered in previous work, as discussed in \autoref{subsubsec:background-microarch-fr}.

\paragraph{Our solution}
To overcome this limitation, we designed a gadget that turns the prefetching mechanism into a distinguisher, bypassing the spatial limitation of Flush+Reload. The gadget works as follows. We probe the cache line \emph{next} to target instructions. In the meantime, we perform a PDA to slow down the conditional instructions (usually used as a probe), enough so that the \emph{next} instructions are prefetched into the cache multiple times. Thus, attackers observe more cache hits if the secret-dependent branch is executed. In our approach, the PDA \emph{creates} the distinguishing behavior that we monitor with Flush+Reload. 

Applied to our case study, our distinguisher is defined as follows: while probing line~\ref{lst:point_decompress_setPoint}, we discernibly obtain more hits when line~\ref{lst:point_decompress_sub} is executed. The reasons are twofold. First, the PDA slows down the execution of the conditional branch by evicting its instructions out of the cache. Second, some CPU optimization gets the probe back in the cache due to the spatial proximity with the conditional branch. 
The repeated reloads cause hits, constituting a characteristic behavior (refer to \autoref{fig:attack_sample}).
As a side note, the number of cache hits may be used to define a confidence coefficient to prune false positives and avoid poisoning our dataset when working with a limited number of traces.

\paragraph{Comparison with the Naive Approach}
The naive approach consists in applying a PDA in order to increase the latency between accesses to the probed instructions. This achieves high temporal resolution while avoiding concurrent access to the probe. However, it was not able to get any result in this context because of the mild instruction discrepancy. 
In our attack, we rely on PDA differently to overcome the spatial limitation of Flush+Reload, while keeping a high temporal resolution. Indeed, we leverage the PDA to slow down the conditional instructions (usually used as a probe) enough to make the prefetcher load \emph{next} instructions into the cache multiple times. The outcome is more observed hits by the attacker. 

\paragraph{Building the gadget}
We experimentally confirmed the efficiency of this approach by distinguishing the mild instruction difference in three different cryptographic libraries. For each one, we built our gadgets the same way. First, we identify the instructions we want to distinguish and get their offset in the shared library. This can be found with basic reverse engineering (\eg using \code{objdump}) of the targetted binary, and enables us to compute the address we will constantly flush. Then, we define the probe on the \emph{next} cache line. In case the probe or the flush target is always executed, the targeted instructions may need to be readjusted (\eg look for a cache line boundary, or different distinguishing instructions). We emphasize that the spatial proximity of the probe is more relevant than its logical proximity, due to the observed prefetcher's behavior.

\subsubsection{Attack Implementation}

\input{resources/tikz/attack_sample}

We target the OpenSSL function \code{ec\_GFp\_simple\_set\_compressed\_coordinates}, represented in \autoref{lst:openssl_impl}. We recall that, in SAE, this function is executed at the end of the password conversion (\autoref{lst:h2c}, line~\ref{lst:h2c_sety}), after going through the 40 iterations of the hunting-and-pecking.

Applying our gadget, we probed the call to the function \code{EC\_POINT\_set\_affine\_coordinates} with Flush+Reload, while performing a PDA on both the call to \code{BN\_usub} and its internal instructions. The PDA aims to increase the latency of \code{BN\_usub} by forcing the execution pipeline to fetch our probe multiple times. On our binary, the monitored instructions are located 101 bytes from the call to \code{BN\_usub}. This means that the probe, and the conditional instructions, are at most one cache line apart, and possibly on adjacent cache lines (considering 64-byte lines). We suspect that such spatial proximity triggers some CPU optimization to fetch in the next instructions. Indeed, we notice that an execution where the parity of \code{pt} and \code{y} is different results in more hits.

In practice, we observed an increase in the number of hits by a factor of five to ten (going from a couple of hits to about 10 hits or more). This difference is easily recognizable, as shown in \autoref{fig:attack_sample}. Here, the threshold (horizontal red line) between a cache-hit and a cache-miss is empirically defined by measuring the average loading latency of each action. We can observe significantly more hits (blue cross under the red threshold) when the call to \code{EC\_POINT\_set\_affine\_coordinates} is executed (\textit{i.e.}, leaking that the parity is not equal). Finally, we attempted to investigate the origin of this behavior caused by some CPU optimization. Our guess was the prefetcher. However, we still obtain the same results even after deactivating the four CPU prefetchers documented in~\cite{IntelArchitectureOptimizationManual} (streamer, Spatial, Data Cache Unit, and Instruction Pointer-based) using Model-Specific Register (MSR) 0x14a.

\subsubsection{Performance and Accuracy}

As with any cache-based attacks, our measurements are susceptible to system noise and CPU optimizations. Therefore, multiple spied observations are required for the same password and MAC addresses to get dependable results. Below, we describe our settings to assess the accuracy of our attack. We conduct our evaluation with 20 different passwords. For each run, we collect 20 traces, while varying one MAC address. This gives a total of 400 samples.

Our findings show that our technique to overcome the spatial limitations of \fr is rewarding. Indeed, by discarding low confidence traces, we could exploit $365/400$ samples, with only 10 miss-predictions, from only two measurements per sample. Three measurements with the same password and MAC addresses are enough to achieve 100\% of usability, and 100\% of accuracy. Our dataset is available in the PoC repository of our attack\footnote{\url{https://gitlab.inria.fr/ddealmei/artifact_dragondoom}}.

Then, we continue our study by computing the average number of hits for three measurements. Henceforth, we refer to this value as the \emph{execution trace}, or simply as a \emph{trace}. As explained previously, for OpenSSL, each trace reveals one bit of information about the password.

\subsection{Comparative~Analysis~of~Previous~Work}
\label{ssec:compare_experimental}

\autoref{tab:comp_previous_works} sums up the average number of measurements (i.e. SAE handshakes spied on) needed to prune all invalid passwords of various dictionaries with high probability ($p>0.95$), as described in \autoref{subsec:dictionary_attack}. We stress that we consider not only the amount of bits leaked by a handshake, but also the number of measurements required to exploit it reliably. This is evaluated by the efficiency criteria described in \autoref{subsec:dictionary_attack}, and computed in the last line of \autoref{tab:comp_previous_works}.
We only include the result of our attack on OpenSSL in our comparative study. Recall that we estimate that other libraries would leak more information, and therefore have a better efficiency coefficient. Hence, \autoref{tab:comp_previous_works} reflects the worst-case scenario for our attack, and we could expect better results on other implementations. Yet, we still outperform~\cite{DBLP:conf/sp/VanhoefR20} and~\cite{DBLP:conf/acsac/BragaFS20}.

With the cache-timing attack presented in~\cite{DBLP:conf/sp/VanhoefR20}, authors learn whether the first iteration of the conversion loop successfully finds an appropriate x-coordinate. They needed to repeat their measurements up to 20 times to get an exploitable leak. The theoretical vulnerability leaks 2 bits on average, but the provided PoC only leaks one bit. \autoref{tab:comp_previous_works} only takes into account their implemented attack, since implementing their theoretical attack may need more measurements. Our results reveal that our attack is about $6.67$ times more efficient.

In~\cite{DBLP:conf/acsac/BragaFS20}, authors leak more information, as they were able to get the exact iteration corresponding to the successful password conversion.
This additional leak allowed attackers to obtain about 2 bits of information on average. However, they still need 10 microarchitectural measurements to get a reliable execution trace. Thus, although our leakage per MAC address is less than~\cite{DBLP:conf/acsac/BragaFS20}, our attack is still more efficient thanks to our precise measurements.

Our work, in addition to exploiting unstudied leakage vector in WPA3, achieves better efficiency: it leaks one bit of information per trace, with only 3 measurements needed for each trace. It is worth noting that this only concerns exploiting hunting-and-pecking in OpenSSL. Since they may leak more bits, better results are obtained for WolfSSL and ell. Moreover, the case of SSWU is quite different for two reasons. First, attacks can only obtain one execution trace, while spying on SAE-PT. Second, a trace may leak more bits with lower probability (refer to \autoref{subsec:bin2bn}). Consequently, this vulnerability is considered less practical, but it still provides important insight into the difficulty of providing secret-independent implementations even for constant-time algorithms by design.

\input{resources/table_compare_previous_work.tex}

%% file: resources/pt_decompress_code.tex
\begin{figure*}
%	\centering
	% Let's make this an lstlisting, not a figure...
	%\makeatletter\def\@captype{lstlisting}\makeatother
	\begin{lrbox}{\verbsavebox}
		\begin{lstlisting}[xrightmargin=.667\linewidth,mathescape=true,
		escapeinside={@}{@},style=codesample,frame=none]
ec_GFp_simple_set_compressed_coordinates(x, pt, ec):
	# Compute y candidate
	y = ...
	# Comply to input format
	if ( BN_is_odd(y) != (pt & 1)):
		if BN_is_zero(y):@\label{lst:openssl_impl-y0}@
			# Handle special case
		BN_usub(y, ec.p, y)
	
	P = EC_POINT_set_affine_coordinates(x, y)	
	return P

BN_bin2bn(buf, n, bn):
	# Skip leading zero's
	for ( ; n > 0 && *buf == 0; buf++, n--):	@\label{lst:openssl_impl_bin2bn-opti}@
		continue
	[...]
	while (n--):
		# read the bytes in
		[...]
	bn_correct_top(ret)
	return ret
@\textcolor{white}{.}@
		\end{lstlisting}
	\end{lrbox}
	\subcaptionbox{\textbf{OpenSSL v1.1.1n}\label{lst:openssl_impl}}{\usebox{\verbsavebox}}\hfill~%
	\vrule~\hfill~%
	\begin{lrbox}{\verbsavebox}
		\begin{lstlisting}[xrightmargin=.366\linewidth,mathescape=true,
		escapeinside={@}{@},style=codesample,frame=none,numbers=none]
wc_ecc_import_point_der_ex(x, ec):
	pt = x[0]
	# Compute y candidate
	y = ...
	# Comply to input format
	if (mp_isodd(y) and pt == 3) or  @\label{lst:wolfssl_impl-if1}@
		(not mp_isodd(y) and pt = 4):@\label{lst:wolfssl_impl-if2}@
		P.y = mp_mod(y, ec.p)               @\label{lst:wolfssl_impl-mod}@
	else:
		P.y = mp_submod(ec.p, y, ec.p)      @\label{lst:wolfssl_impl-sub}@
	return P

mp_read_unsigned_bin(bn, buf, n):
	# Skip leading zero's
	while (c > 0 && b[0] == 0):	@\label{lst:wolfssl_impl_bin2bn-opti}@
		c--; b++
	[...]
	mp_zero (bn)
	while (n-- > 0):
		# read the bytes in
		[...]
	mp_clamp (bn)
	return MP_OKAY
		\end{lstlisting}
	\end{lrbox}
	\subcaptionbox{\textbf{WolfSSL v5.0.0-stable}\label{lst:wolfssl_impl}}{\usebox{\verbsavebox}}~\hfill~%
	\vrule~\hfill~%
	\begin{lrbox}{\verbsavebox}
		\begin{lstlisting}[xrightmargin=.666\linewidth,mathescape=true,
		escapeinside={@}{@},style=codesample,frame=none,numbers=none]
l_ecc_point_from_data(data, pt, ec):
	memcpy(P.x, data, ec.n)
	if pt == 4:
		_ecc_compute_y(ec, P.y, P.x)@\label{lst:ell_impl-seed_even}@
		if (!(P.y[0] & 1)):
			_vli_mod_sub(P.y, ec.p, P.y, ec.p)@\label{lst:ell_impl-y_even}@
	elif pt == 3:
		_ecc_compute_y(ec, P.y, P.x)
		if (P.y[0] & 1):
			_vli_mod_sub(P.y, ec.p, P.y, ec.p)
	return P

_ecc_be2native(bn, buf, n):
	uint64_t tmp[2 * L_ECC_MAX_DIGITS]
	for (i = 0; i < n; i++)
		tmp[n - 1 - i] = l_get_be64(&buf[i])
	memcpy(dest, tmp, n * 8)

@\textcolor{white}{.}@
		\end{lstlisting}
	\end{lrbox}
	\subcaptionbox{\textbf{ell v0.47}\label{lst:ell_impl}}{\usebox{\verbsavebox}}
	\caption{Implementation of the leaking function (\code{set\_compressed\_point\_coordinate} and \code{bin2bn}) in various cryptographic libraries. We extracted only the relevant part and simplified the code to save space. Importantly, \code{pt} represents the point type, aka the compression type, which can be either 3 or 4 for compressed coordinates.}
	\label{lst:practical_implem}
\end{figure*}

%% file: resources/table_affected_projects.tex
\begin{table}
	\begin{threeparttable}
    \begin{tabular}{cl|cccc}
        \multicolumn{1}{l}{}                        &            & \multicolumn{4}{c}{\textbf{Cryptographic library}}               \\
        \multicolumn{1}{l}{}                        &            & \rot{45}{\small OpenSSL}    & \rot{45}{\small WolfSSL} & \rot{45}{\small ell} & \rot{45}{\small CoreCrypto}  \\ \hline
        \multirow{4}{*}{\rot{90}{\textbf{Project}}} & hostap     & $\bullet$            & $\bullet$         &               &                       \\
                                                    & iwd        &                      &                   & $\bullet$     &                       \\
                                                    & FreeRadius\tnote{\dag} & $\bullet$            &                   &               &                       \\ 
                                                    & CoreCrypto &                      &                   &               & $\bullet$             \\ \hline
        \multicolumn{2}{l|}{\textbf{Average leakage \ref{subsec:point_decompression}}}  & 1                 & 1.5           & 2             & 0     \\
        \multicolumn{2}{l|}{\textbf{Average leakage \ref{subsec:bin2bn}}}               & 0.094             & 0.094         & 0             & 0
    \end{tabular}
	\begin{tablenotes}\footnotesize
		\item[\dag] FreeRadius behaves differently than other Wi-Fi daemon, and leaks more data on Vuln. \ref{subsec:bin2bn} (0.131 bits on average)
	\end{tablenotes}				
    \caption{List of the studied SAE implementations, with leveraged cryptographic libraries. 
    Each $\bullet$ means that the implementation supports the library.
    Last lines show the average leakage (in bits) from a single session.}
    \label{table:sae_crypto_lib}
    \end{threeparttable}
\end{table}

%% file: resources/tikz/attack_sample.tex
\begin{figure}
  \begin{subfigure}[b]{0.48\linewidth}
  	\centering
    \begin{tikzpicture}
	    \begin{axis}[
		    width=\linewidth,
		    domain=980:1050,
		    xmin=980, xmax=1050,
			xmajorticks=false,
		    ymin=0,
		    ymax=350,
		    ytick={50, 150, 250},
		    %xtick distance=20,ytick distance=50,
		    xlabel=Time,
		    ylabel=Cycles to reload,
		    enlargelimits=false,
		    axis lines=left,
		]
		    \addplot +[only marks,mark=x,restrict y to domain=0:300] 
		    table[x index=0, y index=2] 
		    {resources/tikz/data/attack_sample-diff_parity-0a44e248f1b4_E2AA67DDF6DA_190807gustavo.dat}; 
		    \addplot [thick,color=red] {120};
	    \end{axis}
    \end{tikzpicture}
    \subcaption{Branch taken}
    \label{fig:attack_sample-diff_parity}
  \end{subfigure}\hfill
  \begin{subfigure}[b]{0.48\linewidth}
  	\centering
  	\begin{tikzpicture}
	  	\begin{axis}[
		  	width=\linewidth,
		  	domain=900:960,
		  	xmin=900, xmax=960,
		  	ticks=none,
		  	ymin=0,
		  	ymax=350,
		  	%xtick distance=20,ytick distance=50,
		  	xlabel=Time,
		  	enlargelimits=false,
		  	axis lines=left,
	  	]
		  	\addplot +[only marks,mark=x,restrict y to domain=0:300] 
		  	table[x index=0, y index=2] 
		  	{resources/tikz/data/attack_sample-same_parity-5a5aa1db29a6_E2AA67DDF6DA_190807gustavo.dat}; 
		  	\addplot [thick,color=red] {120};
  	\end{axis}
  	\end{tikzpicture}
  	\subcaption{Branch not taken}
  	\label{fig:attack_sample-same_parity}
  \end{subfigure}
  \caption{Representation of a measurement when the branch is executed (left) and not executed (right). The red line is the threshold: each blue cross under the line means that the victim has loaded the probe in the cache.}
%   \vspace*{-1em}%These graphs have been acquired by measuring the execution for the password "190807gustavo" with MAC addresses \code{5a5aa1db29a6} on the left, and \code{0a44e248f1b4} on the right.}
  \label{fig:attack_sample}
\end{figure}
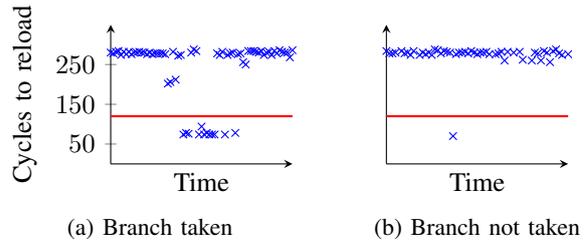

%% file: resources/table_compare_previous_work.tex
\begin{table}
    \centering
\begin{tabular}{l|ccc}
%\cline{2-4}
                                                        & \multicolumn{1}{c}{\textbf{\cite{DBLP:conf/sp/VanhoefR20}}} & \multicolumn{1}{c}{\textbf{\cite{DBLP:conf/acsac/BragaFS20}}} & \multicolumn{1}{c}{\textbf{This work}} \\ \hline
\multicolumn{1}{l|}{rockyou ($1.4\cdot 10^7$)}         & 580                                                          & 160                                                            & \textbf{87}                             \\
\multicolumn{1}{l|}{CrackStation ($3.5\cdot 10^7$)}    & 600                                                          & 170                                                            & \textbf{90}                             \\
\multicolumn{1}{l|}{HaveIBeenPwned ($5.5\cdot 10^8$)}  & 680                                                          & 200                                                            & \textbf{102}                            \\
\multicolumn{1}{l|}{8 characters ($4.6\cdot 10^{14}$)} & 1060                                                         & 320                                                            & \textbf{159}               				\\ \hline
\multicolumn{1}{l|}{Overall efficiency ($\mathcal{D}$)} & 0.05                                    					  & 0.2                                       					   & \textbf{0.33}
\end{tabular}
\caption{Comparison of the number of the measurements needed to prune all wrong passwords of various dictionary for our attack and previous works. Last line shows the overall efficiency of the attack as a ratio between the leakage per MAC address, and the number of required measurements.}
\label{tab:comp_previous_works}
\end{table}

%% file: sections/06_mitigations.tex
%!TEX root = ../2022_dragondoom_to_dragonstar.tex

\subsection{Short Term Mitigation}

The most common way to mitigate this type of vulnerability is to address them at the application-level, with a secret-independent implementation. This approach benefits from two major upsides: (i) the patch is often simple; and (ii) the overhead is minimal and limited to the particular application. It is important to highlight that a peculiar aspect of our vulnerabilities lies in an uncommon interaction between SAE and its cryptographic provider. Indeed, for instance, in all the studied libraries, the function setting a point coordinate never handles the compression format as a secret.

Regarding point decompression, the mitigation is similar in all the affected projects. Both potential point coordinates must be computed, followed by a constant-time selection. After our disclosure, this solution has been implemented by hostap, FreeRadius and iwd. It is worth noting that the original routines \code{set\_compressed\_point\_coordinate} in OpenSSL, WolfSSL and ell still branch on the compression format.

Unlike point decompression, for the BN conversion vulnerability, we included the cryptographic libraries in our disclosure, namely OpenSSL and WolfSSL. This is because such libraries encapsulate the definition of the BN structure. Thus, any related code is reasonably expected to be patched by the concerned libraries. Both OpenSSL and WolfSSL acknowledged that their BN operations are leaky, but refused to patch. Indeed, they argued that it is upon developers' responsibility to avoid calling these functions with secret-dependent values.

This strategy, as convenient as it might seem, is prone to error and only provides relative security (as demonstrated by these vulnerabilities, despite previous mitigations). An implementation bringing formal guarantees regarding the secret independence would provide a more sustainable solution.

\subsection{Dragonstar: Formally Verified Cryptography for Dragonfly}

The root cause of the attacks relies upon the use of cryptographic libraries that are not secret-independent. To address this, we built a plugin for hostap, where all the cryptographic calls within SAE redirect to the HACL* formally verified library. Below, we briefly outline our implementation and its use of HACL*.

\paragraph{hostap structure}

As with most projects of such size, hostap defines multiple abstractions layers. In particular, all cryptographic operations are processed through a common cryptographic API. This enables a very modular approach and eases the support of new libraries. Adding a new library can be done by implementing an interface between the common API and the underlying library for all required functions.

We provided an implementation of the cryptographic API using functions from the HACL* cryptographic library. In particular, we use verified code for the HMAC-SHA256 message authentication code, the NIST P-256 elliptic curve, and the generic BN library. To provide support to the required API, we worked to expose several internal HACL* functions and implement optimizations. In total, we wrap 28 verified functions from HACL* and the code to meet hostap's cryptographic API.

\paragraph{Using verified cryptography from HaCl*}
All the code we use from HACL* is verified for correctness, memory safety, and secret independence. For example, the point decompression function in HACL* is verified to have the F* type given in \autoref{lst:fstar}.

\begin{lstlisting}[
    caption={F* type for the point decompression.}, 
    label={lst:fstar},
    language=fstar,
    basicstyle=\scriptsize,
    frame=lines,
    captionpos=b,
    float,
    floatplacement=H,
]
val decompress: input: lbuffer uint8 33ul -> result: lbuffer uint8 64ul
  -> Stack bool 
  (requires fun h -> live h input /\ live h result /\ disjoint input result)
  (ensures fun h0 success h1 ->
    let compressed = as_seq h0 input in
    let uncompressed = as_seq h1 result in
    (uncompressed,success) == decompress_spec compressed /\
    modifies (loc result) h0 h1)
\end{lstlisting}

This function takes a compressed point (\code{input}) and decompresses it into \code{result}, returning a boolean indicating success or failure.  Both the input and the result are fixed-length arrays (\code{lbuffer}) that are assumed to contain secret bytes, indicated by the type \code{uint8}. By default, we treat all bytes and integers as secrets; if an array is known to contain only public bytes, we would use \code{pub\_uint8} instead of \code{uint8}. Hence, the type given to \code{input} constrains the code of the \code{decompress} function to treat the contents of \code{input} and any value derived from \code{input} as opaque secret values. Branching on the parity of the y-coordinate, for example, would result in a type error, since secret bytes do not have a comparison operation. In other words, the type of \code{input} ensures that the code must be secret-independent with respect to its contents.

In addition to secret independence, the type above also enforces memory safety and correctness. The function is in the \code{Stack} effect, indicating that it only uses the stack and does not allocate or free any memory in the heap.  The precondition (\code{requires}) states that the \code{input} and \code{result} arrays point to valid disjoint locations in the heap. The post-condition (\code{ensures}) says that the output of \code{decompress} matches its spec \code{decompress\_spec} and that the function only modifies the \code{result} array.

Similarly to point decompression, the BN conversion function in HACL* (\code{bn\_from\_bytes\_be}) is verified to be secret independent. In particular, it does not strip leading zeroes and produces a fixed-size BN.

Consequently, by using verified functions from HACL*, we eliminate the two leaks we have explored in this paper, and more generally, we formally guarantee the absence of a large class of timing attacks on our code. Related limitations are discussed in \autoref{sec:discussion}.%\autoref{subsec:remaining_sca}.

\iffalse
\begin{lstlisting}[language=fstar,basicstyle=\footnotesize]
val bn_from_bytes_be:
    #t:limb_t
  -> len:size_t{0 < v len /\ numbytes t * v (blocks len (size (numbytes t))) <= max_size_t}
  -> input:lbuffer uint8 len
  -> result:lbignum t (blocks len (size (numbytes t))) ->
  Stack unit
  (requires fun h -> live h input /\ live h result /\ disjoint input result)
  (ensures  fun h0 _ h1 ->
    as_seq h1 res == Spec.bn_from_bytes_be (v len) (as_seq h0 b) /\
    modifies (loc result) h0 h1)
\end{lstlisting}
\fi

\paragraph{Benchmark}

All benchmarks have been performed using the tool \code{perf} on the same set of inputs while fixing all random values. The setup is the same as described in \autoref{subsubsec:exp-attack-setup}. 
We repeated experiments on 20 different passwords, with fixed MAC addresses and password identifier. Then, we computed the number of cycles required to establish 1,000 sessions with each password. For SAE-PT, we generated the point PT once for each password, and reused the pre-computed value in all subsequent session establishment, as intended by the standard. Thus, repeated sessions smoothed the initial cost of computing PT.

\autoref{fig:benchmark} represents the average number of cycles needed to perform a Dragonfly handshake using hostap for both SAE and SAE-PT. Here, we compare our HACL*-based implementation with OpenSSL, the default cryptographic library in most settings. We also include the OpenSSL build with no assembly code (\textit{i.e.}, noasm).

We highlight two main findings. First, SAE-PT is always faster than SAE. This can be explained by the fact that PT is only computed once. In addition, the mitigations implemented by hunting-and-pecking require to loop over the conversion a fixed number of times, implying dummy iterations. In contrast, SSWU offers a linear workflow with a single conversion, and arithmetic optimizations. Second, HACL* constitutes a good alternative to OpenSSL. Indeed, HACL* is not only formally proved to be secret-independent, but also provides decent efficiency. It outperforms OpenSSL noasm for both SAE and SAE-PT. However, it is slower when assembly code leveraging specialized CPU instructions is used.

\input{resources/tikz/benchmarks.tex}

%% file: resources/tikz/benchmarks.tex
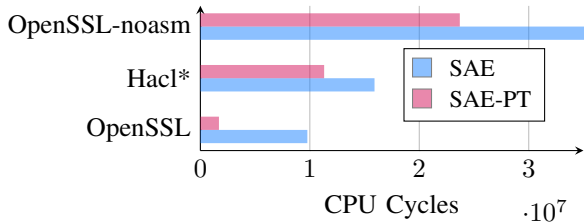
\begin{figure}
%	\centering
	 \begin{tikzpicture}
	  	\begin{axis}[
		  	width = 0.8\linewidth,
		  	height = 3.5cm,
		  	axis lines=left,
		  	xmajorgrids = true,
		  	xlabel = {CPU Cycles},
		  	ytick = data,
		  	xmin=0, xmax=35000000,
			xbar=0pt,
			bar width=5pt,
			tickwidth         = 0pt,
			enlarge y limits  = 0.2,
			enlarge x limits  = 0,
%			symbolic y coords={OpenSSLv1,OpenSSLv3,Dragonstar,OpenSSLv1-noasm,OpenSSLv3-noasm},
			symbolic y coords={OpenSSL,Hacl*,OpenSSL-noasm},
			legend image code/.code={
				\draw [#1] (0cm,-0.1cm) rectangle (0.25cm,0.15cm); 
			},
			legend cell align=left,
		  	legend style={
		  		at={(0.9,0.2)},
		  		anchor=south east,
		  		column sep=1ex,
		  	}
	  	]
	  		% SAE data
		  	\addplot[draw=none,fill=bblue,opacity=0.5] coordinates {%
%		  		(9773908,OpenSSLv1)%
%		  		(10035444,OpenSSLv3)%
		  		(9773908,OpenSSL)%
		  		(15910714,Hacl*)%
		  		(38355528,OpenSSL-noasm)
%		  		(38355528,OpenSSLv1-noasm)%
%		  		(38164825,OpenSSLv3-noasm)%
	  		};
		  	
		  	% SAE-PT data
		  	\addplot[draw=none,fill=rred,opacity=0.5] coordinates {%
%		  		(1696235,OpenSSLv1) %
%		  		(1722293,OpenSSLv3) %
				(1696235,OpenSSL)%
		  		(11308605,Hacl*)%
		  		(23702933,OpenSSL-noasm)
%		  		(23702933,OpenSSLv1-noasm)%
%		  		(23533716,OpenSSLv3-noasm)%
		  	};
%			\addplot[style={blue,fill=blue,mark=none}] coordinates {%
%			  		(OpenSSLv1,18147850)%
%			  		(OpenSSLv3,25198533)%
%			  		(Dragonstar,29390069)%
%			  		(OpenSSLv1-noasm,68107543)%
%			  		(OpenSSLv3-noasm,74518131)%
%		  		};
%			  	
%			  	% SAE-PT data
%			  	\addplot[style={red,fill=red,mark=none}] coordinates {%
%			  		(OpenSSLv1,11765456) %
%			  		(OpenSSLv3,18440713) %
%			  		(Dragonstar,27009494)%
%			  		(OpenSSLv1-noasm,57014407)%
%			  		(OpenSSLv3-noasm,63226925)%
%			  	};
	  	
		  	\legend{{\small SAE}, {\small SAE-PT}}
	  	\end{axis}
	\end{tikzpicture}
	\caption{Performance comparison of our implementation and OpenSSL implementation. All results are obtained by repeating the Dragonfly handshake 1000 times for 20 different passwords. For SAE-PT, we compute PT only once per password, as intended by the specification.}
	% \vspace*{-1em}
	\label{fig:benchmark}
\end{figure}

%% file: sections/07_discussion.tex
%!TEX root = ../2022_dragondoom_to_dragonstar.tex

\noindent\textbf{Is SSWU Worth it?}
SSWU is proposed as a superior alternative in WPA3 both in terms of efficiency and security. Indeed, SSWU is a deterministic mapping, which means that it does not suffer from the inherent secret-dependence issue of a probabilistic approach such as hunting-and-pecking. Moreover, the ongoing standardization process of hash-to-curve functions provides a straightforward and secret-independent implementation of SSWU. Nevertheless, we show that vulnerabilities still sneak into deployed implementations of SSWU. The identified vulnerability is indeed less exploitable in practice, but we still provide important insight regarding the use of third-party libraries. Indeed, although capital, secret-independent design is not enough if the low-level operations (\textit{e.g.}, arithmetic ones) are leaky. Despite our vulnerability, we still believe that SSWU offers better security to settle the cat-and-mouse SCA concerns about Dragonfly.

\noindent\textbf{Is There Any Side-Channel Left?}
WPA3 relies heavily on BN routines that are not secret independent in OpenSSL. This leakage vector has been extensively studied in~\cite{DBLP:conf/uss/WeiserSBS20}. We noticed multiple secret dependence in low-level arithmetic functions, but assessing their exploitability is arduous, which motivated us to provide a secure alternative.
However, as many guarantees as HACL* brings, secret independence does not eliminate all side-channel attacks. In particular, HACL* and therefore, our code may still be vulnerable to fault-injection, or transient execution attacks.
New approaches for verifying code against advanced side-channels are still under active development~\cite{spectre-verification,maskverif}.
For example, the domain-specific language FaCT~\cite{DBLP:conf/pldi/CauligiSJBWRGBJ19} recently evolved to provide guarantees against transient execution attacks~\cite{DBLP:journals/iacr/ShivakumarBBCCG22}.  
If and when mitigations and verification techniques for such attacks get incorporated into HACL*, our implementation will benefit from these defenses automatically. We also note that we do not yet provide a full proof of correctness for Dragonstar with respect to the Dragonfly specification. Instead, only the cryptographic provider is proved secure, which means that other flaws can still sneak into the base code. Our approach does come with an important practical advantage since it allows smooth integration into existing Wi-Fi daemons. 
Indeed, Dragonstar fits the cryptographic-provider structure of hostap, requiring only a minimal change to the existing project. Because the password conversion methods are implemented in the core of hostap, a verified implementation of such routines would have resulted in substantial changes, and integration of generated code which is hard to understand and maintain.

%% file: sections/08_related_work.tex
%!TEX root = ../2022_dragondoom_to_dragonstar.tex

\paragraph{Cache-attacks on cryptographic implementations} In the last decades, numerous works exploited side-channels to break cryptographic implementations. Notably, cache-based side-channel has become a recurrent exploit tool, as shown by the extensive study of Lou \textit{et al.}~\cite{DBLP:journals/csur/LouZJZ21}. They can be classified into two families: \textsc{Prime+Probe}~\cite{DBLP:conf/ccs/Aciicmez07,DBLP:conf/sp/LiuYGHL15,percival2005cache,DBLP:conf/ctrsa/OsvikST06,DBLP:conf/sacrypt/NeveS06,DBLP:conf/ccs/RonenPS18} and Flush+Reload~\cite{DBLP:conf/sp/GullaschBK11,DBLP:conf/uss/YaromF14,DBLP:conf/acsac/AllanBFPY16}.  
In practice, \fr has been applied in various contexts to leak information on RSA private keys~\cite{DBLP:conf/ches/BernsteinBGBHLV17,DBLP:journals/tches/AldayaGTB19,DBLP:conf/uss/GarciaHTGAB20,DBLP:conf/uss/YaromF14} and (EC)DSA nonces~\cite{DBLP:conf/ccs/GarciaBY16,DBLP:conf/uss/GarciaB17,DBLP:conf/ccs/AranhaN0TY20,DBLP:conf/uss/GarciaHTGAB20,DBLP:journals/iacr/YaromB14,DBLP:conf/ches/BengerPSY14,DBLP:conf/ctrsa/PolSY15,DBLP:conf/ccs/ShinKKJH18}. 
In our work, we focus on Dragonfly of WPA3 and show that, similar to any PAKE, it is fragile to any form of leakage due to the low entropy of passwords.

\paragraph{Attacks on PAKE} Recent contributions show a growing interest on the study of PAKE implementations, such as SRP~\cite{DBLP:conf/acns/Russon21,DBLP:conf/ccs/BragaFS21} or more recent design such as OPAQUE~\cite{DBLP:conf/uss/LenGR21}.  
Importantly, after the standardization of Dragonfly in WPA3, several works took a look into SAE implementations and interactions with the network~\cite{DBLP:journals/iet-ifs/ClarkeH14,DBLP:conf/sin/LounisZ19,DBLP:conf/crisis/LounisZ19,DBLP:conf/sp/VanhoefR20,DBLP:conf/acsac/BragaFS20,DBLP:conf/asiapkc/Vanhoef22}.
Of particular interest, they unveiled multiple vulnerabilities including microarchitectural leaks~\cite{DBLP:conf/sp/VanhoefR20,DBLP:conf/acsac/BragaFS20} in different widespread implementations. 
Previous contributions focused on the Wi-Fi daemon layer implementations of Dragonfly, overlooking its interactions with the cryptographic libraries.
In our work, we dig deeper and identify some secret leakage caused by the inner implementations of the cryptographic libraries. Our identified vulnerabilities do not only concern the hunting-and-pecking, but also the recently standardized SSWU. Moreover, we did not only consider proposing some quick patches but also implementing a formally proved secret-independent component that provides all the required cryptographic routines in Dragonfly.
To the best of our knowledge, other symmetric PAKEs, like CPace~\cite{DBLP:journals/tches/HaaseL19}, have not been targeted by such attacks.

\paragraph{Formally verified implementations} Many works have advocated using formal verification to guarantee the correctness and security of cryptographic libraries and protocol implementations (see \cite{cac-sok} for a detailed survey). Fiat-Crypto~\cite{fiat-crypto} and Cryptoline~\cite{cryptoline} can be used to verify field arithmetic (BN) functions in C and assembly for correctness, but not for secret independence. The SAW workbench~\cite{saw-cryptol} and Coq prover~\cite{tweetnacl-coq} have been used to verify full crypto algorithms in C and Java, including selected elliptic curves, for correctness but not for secret independence.  Vale~\cite{vale2} and Jasmin~\cite{jasmin} have been used to verify both the correctness and secret independence of some crypto algorithms in Intel assembly. To our knowledge, HACL* is the only verified crypto library that includes all required algorithms (including HMAC-SHA256 and NIST P-256) and verifies both functional correctness and secret independence. In addition, opting for HACL*, we benefit from its maintainers' experience to recommend the best strategy of build/update in real projects, as they already do for Mozilla and Wireguard.

%% file: sections/09_conclusion.tex
%!TEX root = ../2022_dragondoom_to_dragonstar.tex

We claim that our work is not yet-another-attack against WPA3. 
Indeed, first, we uncover new vulnerabilities in WPA3 implementations that were extensively analyzed either manually or using dedicated tools to check their constant-time nature. This includes the first side-channel attack against the recently deployed SAE-PT.
Second, we provide insight that cryptographic libraries do not consider constant-time code regarding several functions (\textit{e.g.}, compression and bin-to-bn conversion). These functions remained unstudied since they are mostly used with public parameters. Our paper shows that WPA3 calls these functions with secret-dependent values. Third, we explain how to recover the password from the leaked bits. Last, we address this wide attack vector by implementing a provider where all the called functions with secrets are proved secret-independent. In our work, we did not only point out vulnerabilities but also implemented PoCs and conducted experiments to demonstrate the effectiveness of our attack in real-world settings. This involved a novel Flush+Reload gadget to overcome the spatial resolution limitation of the vanilla attack. Our benchmarks show that Dragondoom is both more efficient (requiring fewer side-channel measurements) than previous attacks and applicable to wider settings (including SSWU).
A long-term lesson of our paper is to underline the limitation of former analysis when disregarding interactions with third-party libraries. Formally verifying large implementations is daunting if we require to dig deeper each time an abstraction layer is defined.

Thus, instead of verifying existing implementations, we tackle the issue from the other end and provide Dragonstar, bringing guarantees on the mathematical operations used in Dragonfly for both hunting-and-pecking and SSWU. Our work shows that we can compose the low-level guarantees of HACL* with the sophisticated technical details of WPA3 to obtain a verified cryptographic provider that can be deployed in real-work project as drop-in replacement of OpenSSL. It is true that providing a complete implementation in F* would have offered stronger guarantees, but we argue that our approach of relying on formally-verified modules fits better the need for large modular projects down to the level of code.

%% file: sections/acknowledgment.tex
%!TEX root = ../2022_dragondoom_to_dragonstar.tex

This work benefited from the support of the projects ANR-22-CE39-0005 DRAMA, ANR-18-CE39-0019 MobiS5 and the France 2030 program managed by the French National Research Agency under grant agreement No. ANR-22-PECY-0006.
Daniel De Almeida Braga is funded by the Direction Générale de l'Armement
(Pôle de Recherche CYBER) and a generous gift from Google.

%% file: sections/appendix_vuln_saept.tex
%!TEX root = ../2022_dragondoom_to_dragonstar.tex

Since SAE-PT does not involve a call to the point decompression method, the first vulnerability (\autoref{subsec:point_decompression}) does not apply. However, hostap with both OpenSSL and WolfSSL suffers from the same leakage with the second vulnerability. 

Indeed, hostap calls the \code{bin2bn} routine on six independent secrets along the hash-to-element method: the input of SSWU and both point coordinates output (SSWU is repeated twice, hence the six occurrences). As described in \autoref{subsec:bin2bn}, this would allow attackers to get $k\times 8$ bits of information with probability $6/256^k$. 

Interestingly, the identifier is not attacker-controlled since it is set in a configuration file and not shared on the network, and the SSID cannot be change without the client noticing. Hence, attackers cannot arbitrarily modified public values involved in the computation. This greatly limit the attackers ability to aggregate information by repeating the measurements. The only element they may vary is the curve used for the derivation. 

The standard allows support for up to three different curves, and both WolfSSL and OpenSSL support all of them. Hence, attackers may repeat this measurement up to three times, meaning they could leak 8 bit of information with a probability $p=18/256$, or in 7\% of the cases. This leakage expands to 16 bits in approximately 0.03\% of the cases, which is still relevant considering the wide deployment of WPA3.

On the other hand, ell is not affected since it does not provide the conversion optimization. FreeRadius does not implement SSWU. Experimental results on the \code{bin2bn} conversion methods are presented in \aref{app:results_bin2bn}.

%% file: sections/appendix_res_bin2bn.tex
%!TEX root = ../2022_dragondoom_to_dragonstar.tex

The second vulnerability, affecting both WolfSSL and OpenSSL \code{bin2bn} function, may be exploited using the same Flush+Reload-gadget described in \autoref{ssubsec:fr_gadget}, using the number of cache-hits as an indicator to learn how many leading bytes are skipped.

\begin{figure}[htbp]
    \begin{subfigure}[b]{\linewidth}
        \includegraphics[width=0.95\linewidth]{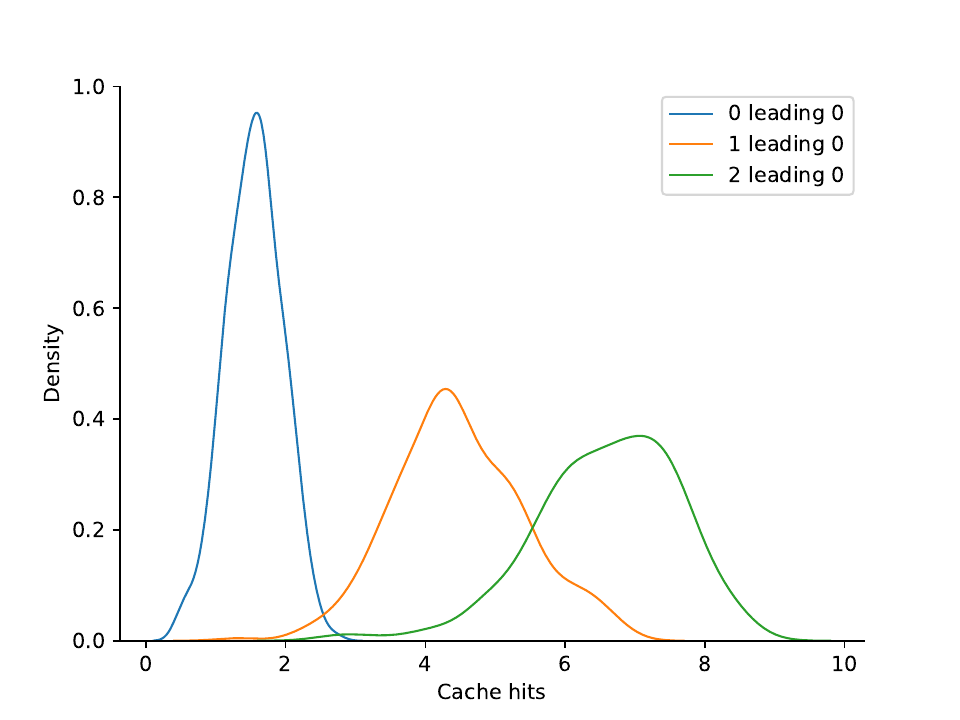}
        \caption{OpenSSL}
    \end{subfigure}
    \begin{subfigure}[b]{\linewidth}
        \includegraphics[width=0.95\linewidth]{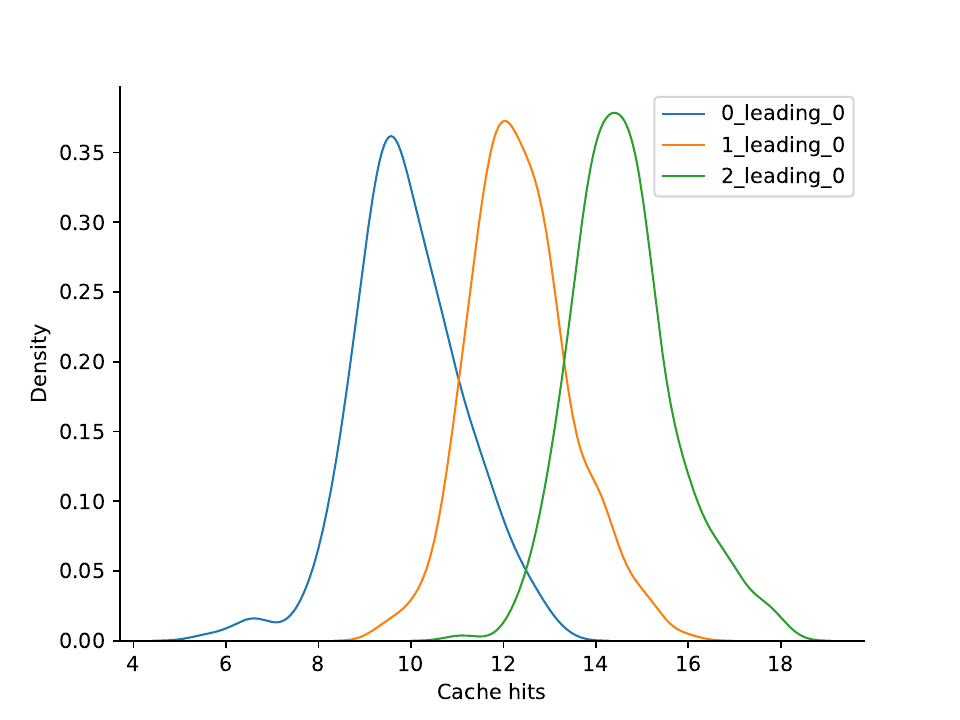}
        \caption{WolfSSL}
    \end{subfigure}
    \caption{Average cache-hits in \code{bin2bn} for different numbers of leading zero.}
    \label{fig:bin2bn_res}
\end{figure}

\autoref{fig:bin2bn_res} represents the average number of hits per measurement, for both OpenSSL and WolfSSL conversion functions. The different curves represent an increasing number of leading zero bytes. We only measured up o two leading zero bytes, as the probability of having more zero leading bytes quickly decreases in our context, as the value is expected to be uniformly distributed. For both implementations, the curves show a clear pattern: the more leading zero bytes, the more hits we observe. 
Each distribution is the result of our attack on 300 samples, executed on 32-byte value to represent a classical use of the function when processing P256-related values.

%% file: sections/appendix_additional_exp.tex
%!TEX root = ../2022_dragondoom_to_dragonstar.tex

For all libraries, we confirmed the leakage in the point \code{set\_compressed\_point\_coordinate} function. The experimentation settings, challenges, solutions and results for OpenSSL are presented in detail in \autoref{ssec:practical_attack}. We followed the same process on WolfSSL and ell, and present our experimental results hereafter. Namely, we face the same challenges regarding the spatial limitation of classical Flush+Reload, and leverage the gadget described in \autoref{ssubsec:fr_gadget} to leak information on the \code{set\_compressed\_point\_coordinate}.

\begin{figure}
    \centering
    \begin{subfigure}[t]{\linewidth}
        \includegraphics[width=0.95\linewidth]{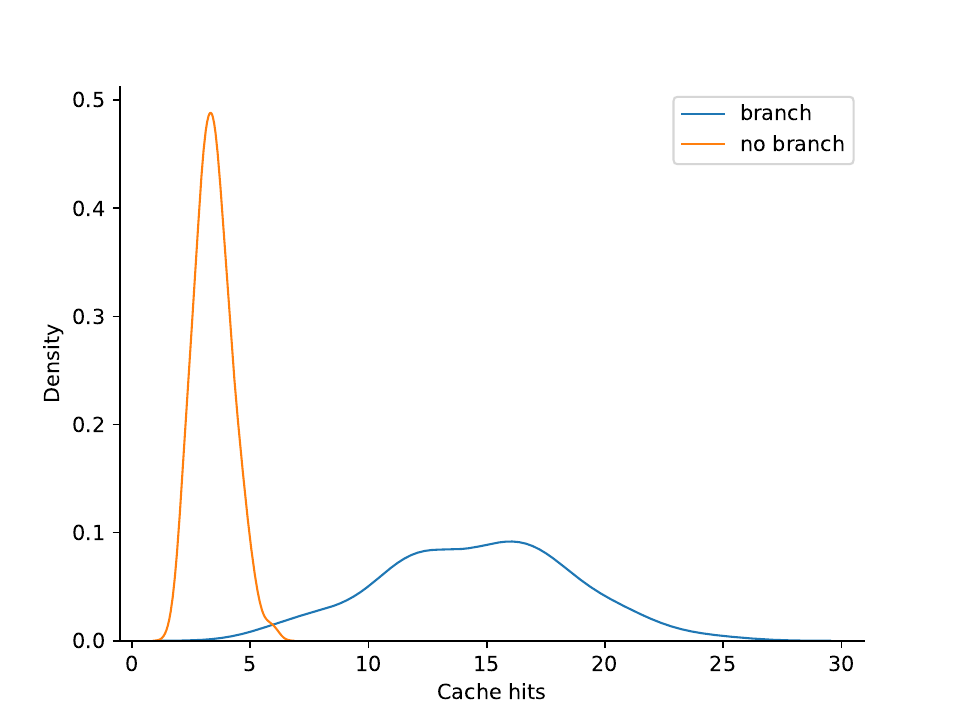}
        \caption{Congruence of y and compression format.}
        \label{fig:exp_ell_parity}
    \end{subfigure}
    \begin{subfigure}[t]{\linewidth}
        \includegraphics[width=0.95\linewidth]{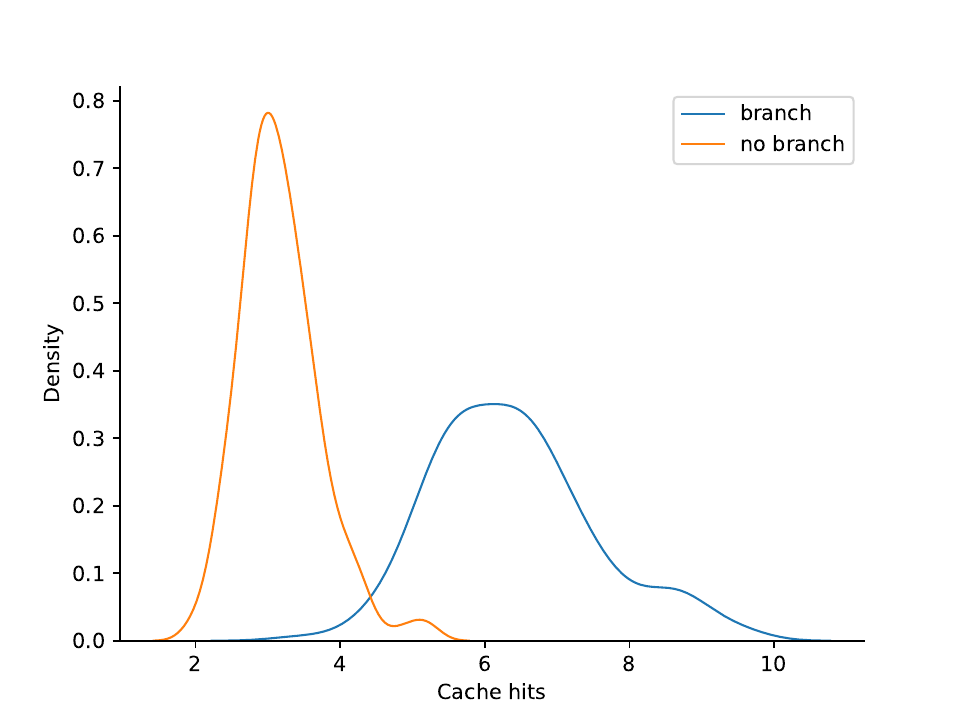}
        \caption{Compression format.}
        \label{fig:exp_ell_fmt}
    \end{subfigure}
    \caption{Density of the number of cache hits in ell for exploiting the two leakages.}
    \label{fig:exp_ell}
\end{figure}

\begin{figure}[htbp]
    \includegraphics[width=0.95\linewidth]{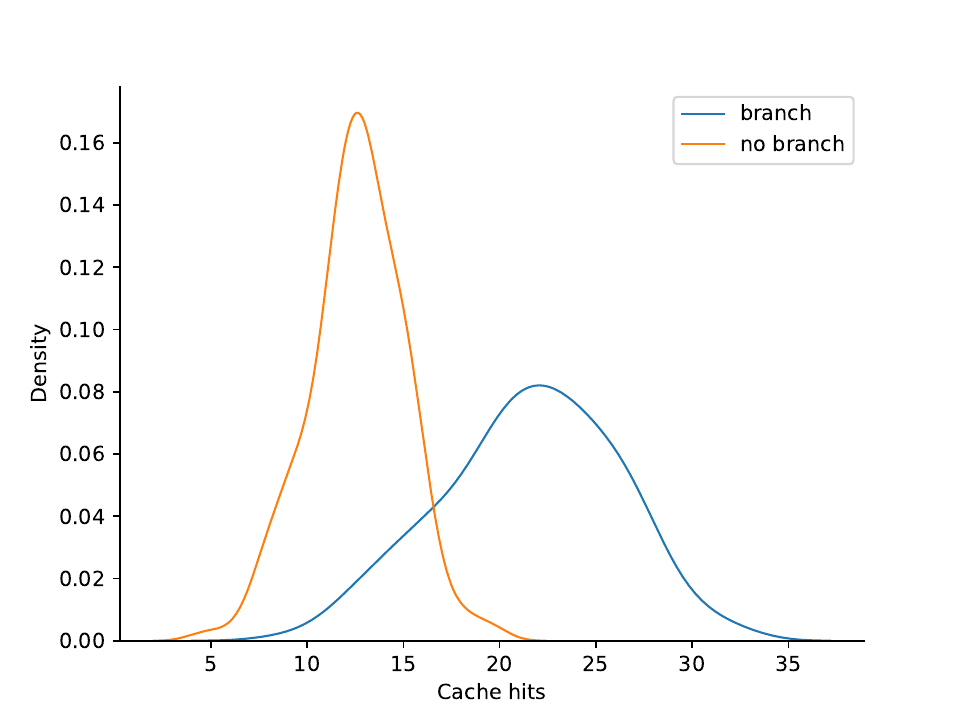}
    \caption{Density of the number of cache hits on the probe when the compression format and the y coordinate have the same parity (blue) or different parity (orange) for WolfSSL.}
    \label{fig:exp_wolfssl}
\end{figure}

In both settings, the libraries were compiled for release, including optimizations. We made measurements with 200 different points resulting in an odd y coordinate and 200 points resulting in an even y coordinate. For each point, we repeated to measurement 5 times and keep the average number of hits to smooth out any noise.
The results are depicted in \autoref{fig:exp_ell} and \autoref{fig:exp_wolfssl}, for ell and WolfSSL, respectively.

\paragraph{ell} For ell, we detailed a more potent leakage, as we can learn both the value of the compression format, and whether it has the same parity as y. Both leakages are caused by branching on the secret values, which we can observe using our gadget. The results are displayed in \autoref{fig:exp_ell}, yielding distinguishable distributions, allowing an attacker to leak the 2 bits of information. 

\paragraph{WolfSSL} We applied our gadget to infer the outcome of the check at line \autoref{lst:wolfssl_impl-if1}-\ref{lst:wolfssl_impl-if2} (\autoref{lst:wolfssl_impl}). Accurately distinguishing the outcome of this branch allows an attacker to learn if the y coordinate and the compression format have the same parity. Both being the outcome of random oracle seeded with password-related value, this leaks 1 bit of information. \autoref{fig:exp_wolfssl} displays two distinct distributions of cache hits when the branch is taken or not. The slight overlapping may be compensated by performing more measurements.